\documentclass[preprints,article,accept,moreauthors,pdftex]{Definitions/mdpi}
\firstpage{1} 
\pubvolume{xx}
\issuenum{1}
\articlenumber{5}
\pubyear{2021}
\copyrightyear{2021}
\history{Received: 22/04/2021, Update: 19/07/2021}

\pdfoutput=1

\usepackage{subcaption}

\Title{Model-predictive control and reinforcement learning in multi-energy system case studies}

\Author{Glenn Ceusters $^{1,2,3,*}$\orcidA{}, Román Cantú Rodríguez $^{4,6}$\orcidB{}, Alberte Bouso García $^{4}$, Rüdiger Franke $^{1}$, Geert Deconinck $^{4,6}$\orcidC{}, Lieve Helsen$^{5,6}$\orcidG{}, Ann Nowé $^{3}$\orcidE{}, Maarten Messagie $^{2}$\orcidF{}, Luis Ramirez Camargo $^{2}$\orcidD{}}

\AuthorNames{Glenn Ceusters, Román Cantú Rodríguez, Alberte Bouso García, Rüdiger Franke, Geert Deconinck, Lieve Helsen, Ann Nowé, Maarten Messagie, Luis Ramirez Camargo}

\address{%
$^{1}$ \quad ABB, Hoge Wei 27, 1930 Zaventem, Belgium; glenn.ceusters@be.abb.com; ruediger.franke@de.abb.com;\\
$^{2}$ \quad Vrije Universiteit Brussel (VUB), ETEC-MOBI, Pleinlaan 2, 1050 Brussels, Belgium; glenn.leo.ceusters@vub.be; Luis.Ramirez.Camargo@vub.be; maarten.messagie@vub.be;\\
$^{3}$ \quad Vrije Universiteit Brussel (VUB), AI-lab, Pleinlaan 2, 1050 Brussels, Belgium; gceusters@ai.vub.ac.be; ann.nowe@ai.vub.ac.be; \\
$^{4}$ \quad KU Leuven, ESAT-ELECTA, Kasteelpark Arenberg 10, 3001 Leuven, Belgium; roman.cantu@kuleuven.be; alberte.bousogarcia@student.kuleuven.be; geert.deconinck@kuleuven.be; \\
$^{5}$ \quad KU Leuven, Dept. of Mech. Engineering (TME), Celestijnenlaan 300, 3001 Leuven, Belgium; lieve.helsen@kuleuven.be; \\
$^{6}$ \quad EnergyVille, 3600 Genk, Belgium.}
\corres{\textbf{Correspondence:} glenn.ceusters@be.abb.com}

\abstract{Model predictive control (MPC) offers an optimal control technique to establish and ensure that the total operation cost of multi-energy systems remains at a minimum while fulfilling all system constraints. However, this method presumes an adequate model of the underlying system dynamics, which is prone to modelling errors and is not necessarily adaptive. This has an associated initial and ongoing project-specific engineering cost. In this paper, we present an on- and off-policy multi-objective reinforcement learning (RL) approach that does not assume a model \textit{a priori}, benchmarking this against a linear MPC (LMPC - to reflect current practice, though non-linear MPC performs better) - both derived from the general optimal control problem, highlighting their differences and similarities. In a simple multi-energy system (MES) configuration case study, we show that a twin delayed deep deterministic policy gradient (TD3) RL agent offers the potential to match and outperform the perfect foresight LMPC benchmark (101.5\%). This while the realistic LMPC, i.e. imperfect predictions, only achieves 98\%. While in a more complex MES system configuration, the RL agent's performance is generally lower (94.6\%), yet still better than the realistic LMPC (88.9\%). In both case studies, the RL agents outperformed the realistic LMPC after a training period of 2 years using quarterly interactions with the environment. We conclude that reinforcement learning is a viable optimal control technique for multi-energy systems given adequate constraint handling and pre-training, to avoid unsafe interactions and long training periods, as is proposed in fundamental future work.}


\usepackage{graphicx}
\usepackage{pdflscape}
\usepackage{longtable}
\usepackage{cleveref}
\usepackage{wrapfig}
\usepackage[ruled,vlined]{algorithm2e}
\usepackage[official]{eurosym}
\usepackage{makecell}
\usepackage{adjustbox}
\begin{document}

\newpage
\section*{Highlights}
\begin{itemize}
    \item An integrated control strategy enables the efficient use of energy at lower costs.
    \item Model-predictive control and reinforcement learning have common mathematical ground.
    \item Reinforcement learning-based energy management doesn’t require \textit{a priori} models
    \item Reinforcement learning can outperform model-predictive control after training.
    \item Safety and fast convergence remain as challenges in reinforcement learning 
\end{itemize}


\section{Introduction}
\par Even today, various forms of energy resources are typically prearranged in a format that solely permits separate operations. However, a number of energy technologies that allow for effective sector coupling are already widely available. This increase in the interconnection between individual energy systems gives rise to the need for integrated control that can further enhance the overall efficiency and performance. Multi-energy, -carrier, -commodity or -utility systems then, in turn, allow for flexibility utilization (i.e. storage, controllable loads) within and across all carriers based on measures such as their energy efficiency, purchasing, emissions, dependability or a combination thereof. For example, combined heat and power plants (CHP) that simultaneously produce heat and electricity via the use of natural gas, virtually linking electricity, heating, and natural gas systems. While, on the other hand, heat pumps generate heat by using electricity and, thereby, link electricity and heating systems. Even variable and fixed-displacement pumps are the linking technology between electrical and water or hydraulic systems. Compressors link air with electricity, and so forth.

\par To ensure the optimum level of operation of such multi-energy systems, specific set-points are required to establish and maintain the total operating costs at a minimum while fulfilling all system constraints. As flexibility utilization exhibits dynamic behaviour and dependency between successive time steps, optimisation across numerous time steps is necessary. Additionally, significant forecast uncertainties need to be managed so that the stability of the multi-energy system is preserved. Model-predictive-control (MPC) offers a suitable control strategy that takes into consideration both system dynamics (i.e. variation in demand, pricing and environment) and when formulated as a stochastic finite-horizon control problem, forecast uncertainties.

\par
\begin{figure}[!htbp]
\centering
\begin{subfigure}{.5\textwidth}
  \centering
  \includegraphics[width=1\linewidth]{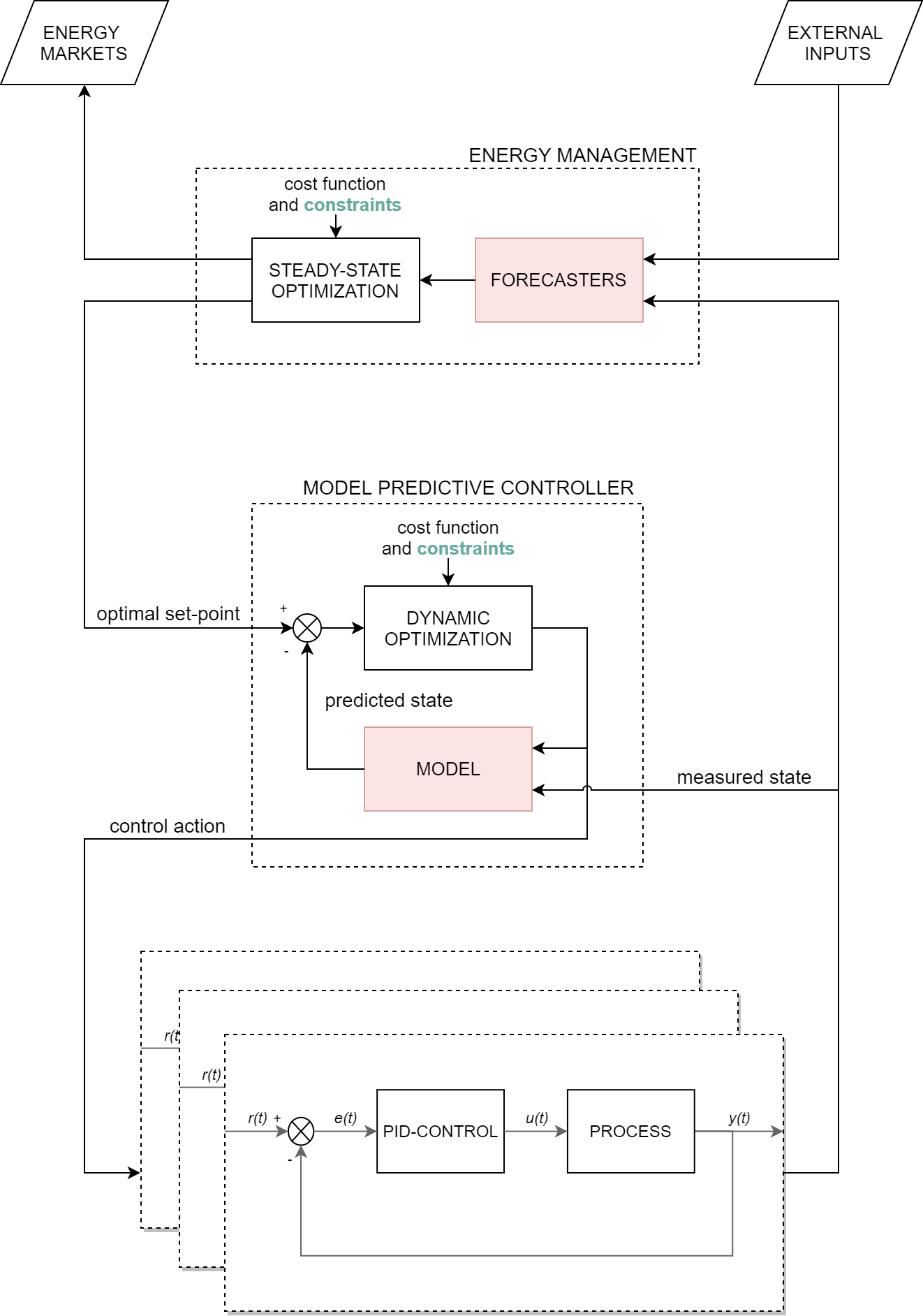}
  \caption{model-predictive control}
  \label{fig:mpc}
\end{subfigure}%
\begin{subfigure}{.5\textwidth}
  \centering
  \includegraphics[width=1\linewidth]{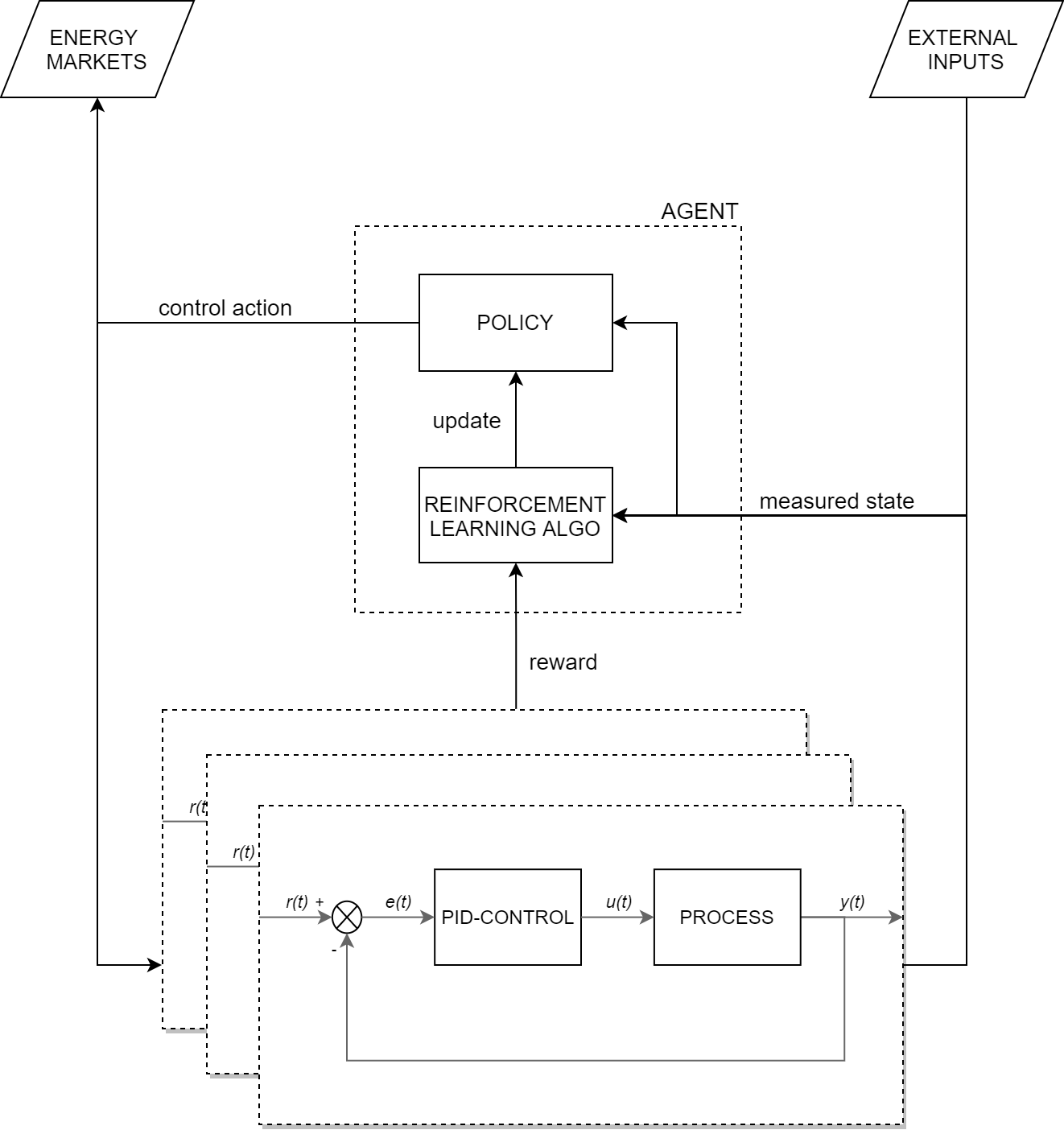}
  \caption{reinforcement learning}
  \label{fig:rl}
\end{subfigure}
\caption{block diagrams comparison in unified terminology. }
\label{fig:blockdiagrams}
\end{figure}

 However, a model-predictive controller presumes an adequate model of the technologies involved (see \autoref{fig:mpc}). This model is required to generate future trajectories that are used during the optimisation process. In cases where such a model is not available or economically viable, MPC will not be able to achieve optimal control. Even if a model is present, two undesirable characteristics reduce its real-world usage. First, should the human modeller make a mistake - indicated in red on \autoref{fig:mpc} - or the model itself is inaccurate, MPC will always operate sub-optimally. Second, regardless of whether the original model is suitable, it may not consistently represent the actual system characteristics; e.g., ageing effects or system modifications, the MPC will continue with its optimisation irrespective of any changes to the system dynamics/structure. The absence of an adaptive configuration in the MPC can result in undesirable outcomes (sub-optimal performance) and, in the extreme case, be responsible for the loss of user comfort or system damage (e.g. turbine control). Furthermore, due to the finite-horizon formulation of an MPC (see \autoref{equation13a}), long-term constraints and objectives are challenging to consider adequately (e.g., the energy balance of geothermal wells, seasonal storage, peak load taxation). Current developments in MPC tackle the next three areas of opportunity: 1) the use of automated toolchains to generate accurate non-linear white-box models, 2) the increase of MPCs' adaptivity and 3) the addition of a shadow-cost to the objective function to take into account long-term effects in the system \cite{Jorissen2021NMPC, CupeiroFigueroaIago2020AMfL}.

\par While multiple architecture variations exist, the unconstrained error handling is performed by classic proportional–integral–derivative (PID) controllers (see \autoref{fig:blockdiagrams}). The MPC consists of a steady-state optimisation problem over a prediction horizon and a dynamic optimisation problem over a real-time control horizon (see \autoref{fig:mpc}). In contrast, a model-free reinforcement learning (RL) agent learns the dynamics of the system and a mapping between states and actions (i.e. the policy, see \autoref{fig:rl}) from direct interaction with the environment or (in some cases) from collected historical data. This is accomplished without the need for prior knowledge of the system.

\par Note that \autoref{fig:blockdiagrams} uses unified terminology, as in the RL community they talk about actions, states and observations without specifying that these are \textit{control} actions and \textit{measured} states (which can imply noisy measurements). 

\subsection{Contribution and outline}
\par Model-predictive control and multi-objective reinforcement learning have never been, to the best of our knowledge, benchmarked simultaneously in a multi-energy management systems context. This allows for the demonstration of the mathematical synergy between both approaches and the formulation of fundamental future work.

\par In \autoref{chapter2} a state-of-the-art summary is presented, \autoref{chapter3} contains preliminaries regarding the optimal control problem, while in \autoref{chapter4} the problem towards an energy management system is formulated. Thereafter, \autoref{chapter5} describes the methodology and multi-energy systems case studies. Finally, \autoref{chapter7} discusses the results and \autoref{chapter8} presents the conclusions and future work. 

\section{State-of-the-art} \label{chapter2}
\subsection{Discussion}

\par Functioning as a starting point, the \textit{energy hub} concept was presented and formulated by Geidl and Andersson, where they provided a model intended for the optimal power dispatch problem in systems with multiple energy carriers \cite{Geidl2005OptimalCarriers}. Through introducing the matrix model and dispatch factors intended for obtaining the contribution of every converter coming from diverse inputs and matching various outputs (MIMO), these authors published an optimisation framework intended for the earlier mentioned optimal power dispatch problem. The same authors \cite{Geidl2005AFlow}, have additionally formulated the optimal power flow (OPF) problem among multiple distinct energy infrastructures, including natural gas, electricity and district heating networks in the context concerning interconnected energy hubs pertaining to multi-carrier energy systems. Furthermore, in \cite{Geidl2005OperationalSystems} they also presented a design optimisation approach to acquire the optimal coupling matrix depending on a specified demand and objective function. These models had been provided in steady-state condition and time-dependent parameters were not taken into consideration. Thereafter,  \cite{Geidl2007IntegratedSystems} wraps up providing a general framework for modelling multi-carrier energy systems and their associated optimisation problems (optimal dispatch, power flow, and design). They also included energy storage systems in the model, taking the time dependency into consideration, which has been discussed in several practical examples.

\par Following the formulation of the energy hub concept, \cite{Arnold2009Model-basedSystems} proposed a centralised model predictive controller. The foregoing controller considers the present dynamics and constraints, and moreover adapts to anticipated - read, forecasted - variations of load and energy prices. Simulations are also presented wherein the proposed scheme is applied to a three interconnected hubs benchmark system. Furthermore, the performance of several prediction horizons of varying lengths has been compared. While the total operation cost decreases with an increasing length of the prediction horizon, the computational effort increases accordingly. For solving the overall optimisation problem in a distributed manner (in an attempt to overcome this computational burden), a distributed model predictive control approach was proposed \cite{arnold2010distributed}. Within a simulated case study, they have analysed the performance of intermediate solutions acquired through the iterations using the proposed approach to make certain that applying the control strategy to a real system provides feasible solutions. Where further, a cooperative behaviour was identified, in which the neighbouring agents assist to the system-wide objective. Thereafter, Arnold et al. \cite{arnold2011predictive} demonstrated that this approach could also be effective in situations in which prediction uncertainties arise regarding the in-feed and demand profiles of renewable energy systems. According to the outputs of the simulations, the main aspect of prediction uncertainties can be compensated for through the use of storage devices as opposed to utilising balancing energy or backup generators.
\par While not explicitly mentioned here, but summarised in \autoref{table: SoA summary}, this framework has seen multiple extensions towards a general unit commitment, demand-side management, optimal-power flow (non-)linear model predictive controller for the multi-energy management problem participating in various energy and grid balancing markets.

\par With the motivation to avoid \textit{a priori} modelling of the underlying system dynamics (as required in a model-predictive controller), a model-free multi-agent reinforcement learning approach was given \cite{Bollinger2016Multi-agentSystems} – where \cite{Gazafroudi2017ASystems} presented a general review of multi-agent energy management systems and \cite{marinescu2017prediction} of multi-agent reinforcement learning. \cite{Rayati2015ApplyingGrid, Ye2020Model-FreeLearning, Wang2020EnergyLearning} also proposed a model-free reinforcement learning, namely Q-learning and deterministic policy gradient, multi-energy management system. The same authors also successfully applied reinforcement learning to the optimal design problem of energy hubs in \cite{Rayati2015ApplyingGrid} and as demand-side management (DSM) system in \cite{Sheikhi2016DemandSystems}.

\par Taking a step back to single-energy systems' literature, \cite{Vanhoudt2015H2020Update, Vanhoudt2017StatusProject, Claessens2018Model-freeNetwork} presented similar motivation for the STORM controller for district heating systems, where a model-free reinforcement learner was proposed in contrast to model-predictive-control, e.g. \cite{verrilli2016model, Lie-Jensen2018ModelSystem}. \cite{Claessens2018Model-freeNetwork} even pointed out that, as discussed in \cite{ernst2008reinforcement}, incorporating basic domain knowledge along with a model-free approach is estimated to lead to an improved performance in decreased learning times. This domain knowledge could be integrated through information about the shape of the policy \cite{busoniu2010reinforcement} as well as through utilizing a model, as in \cite{Lampe2014ApproximateQ-Iteration}. This reinforcement learning trend, for the formulated problem, is not only found in the thermal energy sector (at the building level, e.g. \cite{Kazmi2017SmartOpportunities}), but also in the electrical energy sector, e.g. as reviewed in \cite{Glavic2017ReinforcementPerspectives}, and formally presented in the unit commitment \cite{Dalal2015ReinforcementProblem} and in the economic dispatch problem \cite{Abouheaf2014ApproximateProblems}.

\subsection{Conclusion of state-of-the-art}

\par The array of studies previously summarised (see \autoref{table: SoA summary}) reveals that there is a vast variety in scope, scale, objective and technique involved in the optimisation and control of multi-energy systems. While the scope ranges from design, distribution and operation, the scale varies from building to district. The specified objective is generally cost-driven, being the total cost of ownership (TCO) – hence both capital and operational expenses – for design purposes and running costs, with or without maintenance, for operational objectives. The latter is furthermore dependent on the considered optimisation problem and is only seldom considered simultaneously (e.g. two-level unit commitment and economic dispatch), while considered separately has been practically implemented and validated many times.

\par Consensus was found with respect to a distributed control architecture over a centralised approach to preserve the scalability required in the multi-energy management system (read \autoref{chapter4}). Furthermore, linear and integer programming techniques, are considered the current mature state-of-the-art for solving adequately presumed system models. On the other hand, reinforcement learning, has recently seen considerable recognition by several authors, including the multi-energy research community. It brings and delivers the promise of high generalisability and adaptability at low initial and ongoing cost (i.e. human modelling effort) while considering stochastic environments over an infinite-horizon reward stream, given a sufficient training budget, i.e. time to interact with its environment and/or an adequate pre-training process.

\par In contrast, supervised learning algorithms (i.e., used as data-driven plant models in an MPC) have no data exploration step, thus, even in large datasets, the underlying system dynamics may not be adequately learned (e.g. when the system is always operational, the idle state and its transition are unknown). This may result in the loss of comfort or, in extreme cases, in system failure. Furthermore, reinforcement learning has immature stability, feasibility, robustness theory, and constraint handling \cite{Gorges2017RelationsLearning}, as typical exploration strategies rely on some randomness in the selection of control actions.

\par To conclude this section, a comparison of MPC and RL is shown in \autoref{tab: MPC&RLcomparison}. It shows the main advantages and disadvantages that each technique has with respect to each other according to their current state-of-the-art. 

\begin{table}[!hb]
    \centering
    \begin{adjustbox}{width=1\textwidth}
    \begin{tabular}{l|l|l}
    \rowcolor[HTML]{efefef} 
    \textbf{Characteristic} & \textbf{Model Predictive Control} & \textbf{Reinforcement Learning}\\
    \hline
    Model requirement (a priori) & Yes & \textbf{No} \\
    Model convexity & Usually required & \textbf{Not required} \\
    Adaptivity & Immature (Usually based on robustness) & \textbf{Mature (inherent)} \\
    Online complexity & High (except explicit and neural MPC) & \textbf{Low} \\
    Offline complexity & \textbf{Low (except explicit and neural MPC)} & High \\
    Stability theory & \textbf{Mature (e.g. based on terminal cost)} & Immature \\
    Feasibility theory & \textbf{Mature (e.g. based on terminal constraints)} & Immature \\
    Robustness theory & \textbf{Mature (e.g. based on tubes or ISS)} & Immature \\
    Constraint handling & \textbf{Mature (inherent)} & Immature
    \end{tabular}
    \end{adjustbox}
    \caption{Comparison between characteristics of MPC and RL, adapted from \cite{Gorges2017RelationsLearning}.}
    \label{tab: MPC&RLcomparison}
\end{table}
Note that \textit{safety}, as mentioned in \autoref{table: SoA summary}, is considered to be the grouping of stability theory and constraint handling (i.e., if these properties are fulfilled, one can safely apply the technique in the real world).

\newpage

\begin{landscape}
	\begin{center}
	\begin{longtable}{p{5.1cm}|c|c|c|c|c|c|c|c}
		\caption{State-of-the-art summary. MES = multi-energy system; DMES = distributed multi-energy system; SES = single-energy system; UC = unit commitment; OPF = optimal power flow; DSM = demand-side management; DAO = day-ahead optimisation; IDO = intra-day optimisation; RTO = real-time optimisation; TOU = time-of-use; MPC = model-predictive control; NMPC = non-linear model-predictive control; RL = reinforcement learning; IP= interior-point method; sim = simulation.} \\\rowcolor[HTML]{efefef}
		\label{table: SoA summary}
		\textbf{Reference} & \textbf{Problem} & \textbf{Energy system} & \textbf{Goal} & \textbf{Market} & \textbf{Approach} & \textbf{Architecture} & \textbf{Safety\footnotemark} & \textbf{Environment} \\\hline
		\raggedright (\citeauthor{Geidl2005OptimalCarriers}, \citeyear{Geidl2005OptimalCarriers})\cite{Geidl2005OptimalCarriers} & model & MES & UC & n.a. & n.a. & n.a. & n.a. & sim \\\hline
		\raggedright (\citeauthor{Geidl2005AFlow}, \citeyear{Geidl2005AFlow})\cite{Geidl2005AFlow} & model &	MES &	OPF &	n.a. &	n.a. &	n.a. &	n.a. &	sim \\\hline
		\raggedright (\citeauthor{Geidl2005OperationalSystems}, \citeyear{Geidl2005OperationalSystems})\cite{Geidl2005OperationalSystems} & design &	MES &	TCO &	n.a. &	n.a. &	n.a. &	n.a. &	sim \\\hline
		(\citeauthor{Arnold2009Model-basedSystems}, \citeyear{Arnold2009Model-basedSystems})\cite{Arnold2009Model-basedSystems} &	control &	MES &	UC &	DAO &	MPC &	centralised &	yes &	sim \\\hline
		(\citeauthor{arnold2010distributed}, \citeyear{arnold2010distributed})\cite{arnold2010distributed} &	control &	MES &	UC &	DAO &	MPC &	distributed &	yes &	sim \\\hline
		(\citeauthor{ramirez2013optimal}, \citeyear{ramirez2013optimal})\cite{ramirez2013optimal} & control &	MES &	UC &	DAO, RTO &	MPC &	centralised &	yes &	sim \\\hline
		(\citeauthor{ramirez2015scheduling}, \citeyear{ramirez2015scheduling})\cite{ramirez2015scheduling} & control &	MES &	UC &	DAO, RTO &	MPC &	centralised &	yes &	lab \\\hline
		(\citeauthor{Xu2015HierarchicalSystems}, \citeyear{Xu2015HierarchicalSystems})\cite{Xu2015HierarchicalSystems} &	control & MES &	UC, OPF &	DAO, RTO &	MPC &	centralised &	yes &	sim \\\hline
		\raggedright (\citeauthor{skarvelis2014agent}, \citeyear{skarvelis2014agent})\cite{skarvelis2014agent} & control &	MES &	UC &	DAO &	MPC &	distributed &	yes &	sim \\\hline
		\raggedright (\citeauthor{Skarvelis-Kazakos2016MultipleAgents}, \citeyear{Skarvelis-Kazakos2016MultipleAgents})\cite{Skarvelis-Kazakos2016MultipleAgents} & control &	MES &	UC &	DAO &	MPC &	distributed &	yes &	lab \\\hline
		(\citeauthor{batic2014integrated}, \citeyear{batic2014integrated})\cite{batic2014integrated} &	control &	MES &	UC, DSM &	DAO &	MPC &	centralised &	yes &	sim \\\hline
		(\citeauthor{Batic2016CombinedBuildings}, \citeyear{Batic2016CombinedBuildings})\cite{Batic2016CombinedBuildings} & control &	MES &	UC, DSM &	DAO &	MPC &	centralised &	yes &	lab \\\hline
		(\citeauthor{Maurer2016OptimalGrids}, \citeyear{Maurer2016OptimalGrids})\cite{Maurer2016OptimalGrids} & control &	MES &	UC &	DAO &	NMPC &	centralised &	yes &	sim \\\hline
		(\citeauthor{Booij2013Multi-agentDistricts}, \citeyear{Booij2013Multi-agentDistricts})\cite{Booij2013Multi-agentDistricts} &	control &	MES &	UC, DSM &	RTO &	n.a. &	distributed &	n.a. &	sim \\\hline
		(\citeauthor{Blaauwbroek2015OptimalSystem}, \citeyear{Blaauwbroek2015OptimalSystem})\cite{Blaauwbroek2015OptimalSystem} & control &	MES &	UC, DSM &	DAO, RTO &	MPC	& centralised &	yes &	sim \\\hline
		(\citeauthor{Blaauwbroek2015DecentralizedSystems}, \citeyear{Blaauwbroek2015DecentralizedSystems})\cite{Blaauwbroek2015DecentralizedSystems} & control &	MES &	UC, DSM &	DAO, RTO &	MPC &	distributed &	yes &	sim \\\hline
		(\citeauthor{Shi2016EnergyApproach}, \citeyear{Shi2016EnergyApproach})\cite{Shi2016EnergyApproach} & control &	MES &	UC, DSM &	DAO, RTO &	MPC &	distributed	& yes &	sim \\\hline
		\raggedright (\citeauthor{cesena2016operational}, \citeyear{cesena2016operational})\cite{cesena2016operational} & control &	DMES &	UC, OPF &	DAO	& MPC &	centralised &	yes &	sim \\\hline
		(\citeauthor{Bollinger2016Multi-agentSystems}, \citeyear{Bollinger2016Multi-agentSystems})\cite{Bollinger2016Multi-agentSystems} & design &	DMES &	UC &	DAO	& RL &	distributed &	no &	sim \\\hline
		(\citeauthor{Vanhoudt2015H2020Update}, \citeyear{Vanhoudt2015H2020Update}, \citeyear{Vanhoudt2017StatusProject}; \citeauthor{Claessens2018Model-freeNetwork}, \citeyear{Claessens2018Model-freeNetwork})\cite{Vanhoudt2015H2020Update, Vanhoudt2017StatusProject, Claessens2018Model-freeNetwork} &	control &	SES &	DSM &	n.a. &	RL &	distributed &	no &	sim \\\hline
		(\citeauthor{Kazmi2016GeneralizableNZEB}, \citeyear{Kazmi2016GeneralizableNZEB}; \citeauthor{Kazmi2016DemonstratingBuildings}, \citeyear{Kazmi2016DemonstratingBuildings})\cite{Kazmi2016GeneralizableNZEB, Kazmi2016DemonstratingBuildings} & control &	SES &	UC &	DAO &	RL &	centralised &	no &	sim \\\hline
		(\citeauthor{Kazmi2017SmartOpportunities}, \citeyear{Kazmi2017SmartOpportunities})\cite{Kazmi2017SmartOpportunities} &	control &	SES &	UC &	DAO &	RL &	centralised &	no &	sim \\\hline
		(\citeauthor{Kazmi2018Gigawatt-hourSystems}, \citeyear{Kazmi2018Gigawatt-hourSystems})\cite{Kazmi2018Gigawatt-hourSystems} &	control &	SES &	UC &	DAO &	RL &	centralised &	no &	lab \\\hline
		(\citeauthor{Abouheaf2014ApproximateProblems}, \citeyear{Abouheaf2014ApproximateProblems})\cite{Abouheaf2014ApproximateProblems} & control &	SES &	UC &	RTO &	RL &	centralised &	no &	sim \\\hline
		(\citeauthor{Dalal2015ReinforcementProblem}, \citeyear{Dalal2015ReinforcementProblem})\cite{Dalal2015ReinforcementProblem} & control &	SES &	UC &	DAO &	RL &	centralised &	no &	sim \\\hline
		(\citeauthor{Nagy2017InvestigatingControl}, \citeyear{Nagy2017InvestigatingControl})\cite{Nagy2017InvestigatingControl} &	control &	SES &	UC &	DAO &	RL &	centralised &	no &	sim \\\hline
		(\citeauthor{Tomin2019DeepSources}, \citeyear{Tomin2019DeepSources})\cite{Tomin2019DeepSources} &	control &	SES &	UC, DSM &	n.a. &	RL &	centralised &	no &	sim \\\hline
		(\citeauthor{Latifi2020AMicro/Nano-Grids}, \citeyear{Latifi2020AMicro/Nano-Grids})\cite{Latifi2020AMicro/Nano-Grids} &	control &	SES &	DSM &	n.a. &	RL &	distributed &	no &	sim \\\hline
		(\citeauthor{Rayati2015ApplyingGrid}, \citeyear{Rayati2015ApplyingGrid})\cite{Rayati2015ApplyingGrid} & control &	MES &	UC &	DAO &	RL &	centralised &	no &	sim \\\hline
		(\citeauthor{Sheikhi2016DemandSystems}, \citeyear{Sheikhi2016DemandSystems})\cite{Sheikhi2016DemandSystems} & control &	MES &	DSM &	DAO &	RL &	centralised &	no &	sim \\\hline
		(\citeauthor{Mbuwir2018BatteryLearning}, \citeyear{Mbuwir2018BatteryLearning})\cite{Mbuwir2018BatteryLearning} &	control &	MES &	UC &	DAO &   RL &	centralised &	no &	sim \\\hline
		(\citeauthor{Wang2019Bi-levelSystem}, \citeyear{Wang2019Bi-levelSystem})\cite{Wang2019Bi-levelSystem} &	control &	MES &	UC &	TOU &   RL + IP &	distributed &	yes &	sim \\\hline
		(\citeauthor{Ahrarinouri2020Multi-AgentBuildings}, \citeyear{Ahrarinouri2020Multi-AgentBuildings})\cite{Ahrarinouri2020Multi-AgentBuildings} &	control &	MES &	UC, DSM &	DAO &   RL &	distributed &	no &	sim \\\hline
		(\citeauthor{Ye2020Model-FreeLearning}, \citeyear{Ye2020Model-FreeLearning})\cite{Ye2020Model-FreeLearning} &	control &	MES &	DSM &	TOU &   RL &	centralised &	no &	sim \\\hline
		(\citeauthor{berkenkamp2017safe}, \citeyear{berkenkamp2017safe})\cite{berkenkamp2017safe} &	robotics &	n.a. &	n.a. &	n.a. &	RL &	centralised &	yes &	sim \\\hline
		\textbf{This work} & \textbf{control} &	\textbf{MES} &	\textbf{UC} &	\textbf{DAO} &	\textbf{MPC, RL} &	\textbf{centralised} &	\textbf{no} &	\textbf{sim} \\\hline
	\end{longtable}
	\end{center}
\footnotetext{Assurance of stable and reliable operation.}
\end{landscape}
\newpage

\section{Preliminary - Optimal control problem} \label{chapter3}
\par To highlight the differences and similarities of the main techniques presented in this paper, being MPC and RL, we start our formulation from the general optimal control problem. As (multi-) energy systems are continuous in nature, this is also our starting point - underlying the common discrete-time assumption.

\subsection{Continuous system}
\par Consider a continuous time-varying stochastic system in the form of:

\begin{equation}\label{equation1}
    dx=f(t,\ x_t,\ u_t)dt+dW_t
\end{equation}
where \(x_t\) is the \textit{n}-dimensional \textit{state} vector, \(u_t\) the \textit{m}-dimensional \textit{action} or \textit{control} vector and \(W_t\) a Wiener process, i.e., some stochastic noise. The problem is to find a control signal \( u(t_i\rightarrow t_f) \) so that the cost function

\begin{equation}\label{equation2}
    C(t_i,\ x_i,\ u(t_i\rightarrow t_f))=E \bigg\{ L_T(x_{t_f})+\int_{t_i}^{t_f} L(t,x_t,u_t) dt \bigg\}
\end{equation}
is minimal, where in general terms \( L(t,x_t,u_t) \) is the stage \textit{loss} and \( L_T(x_{t_f}) \) the terminal loss. However, given the stochastic nature of the considered system (\autoref{equation1}), one can only hope to minimize the \textit{expectation} \( E\{ \cdot \} \) of this cost function over all stochastic trajectories that start in \(x_i\). This results in the objective function:

\begin{equation}
    J(t,\ x_t)= \min_{u(t\rightarrow t+dt)} \Bigg( E \bigg\{\int_{t}^{t+dt} L(t,x_t,u_t)dt + J(t+dt,\ x_{t+dt}) \bigg\} \Bigg)\ , \ \forall \ x_t
\end{equation}
However, these equations only hold for problems with a resulting end state at time \( t_f \). The \textit{discounted} infinite-horizon version of \autoref{equation2} is then:

\begin{align}
    C(t_i,\ x_i,\ u(t_i\rightarrow \infty)) &= \lim_{t_f \rightarrow \infty} E \bigg\{ L_T(x_{t_f})+\int_{t_i}^{t_f} L(t,x_t,u_t) dt \bigg\} \\[1em]
    C(t_i,\ x_i,\ u_t) &= E \bigg\{\int_{t_i}^{\infty} \gamma^t L(t,x_t,u_t) dt \bigg\}
\end{align}
where \( 0 < \gamma^t \leq 1 \) is the discount factor, so that the objective function becomes:

\begin{equation}
    J_\infty(x_t)= \min_{u_t \ \in \ U_t} \Bigg( E \bigg\{\int_{t}^{\infty} \gamma^t L(t,x_t,u_t)dt \bigg\} \Bigg)\ , \ \forall \ x_t
\end{equation}

\subsection{Discrete-time approximation}
\par We now consider a discrete time-varying stochastic system in the form of:

\begin{equation}
    x_{t+1}=f(t,\ x_t,\ u_t,\ w_t)
\end{equation}
note that with this formulation, \( w_t \) can enter the dynamics in any form, resulting in the cost function:

\begin{equation}\label{equation8}
    C(t_i,\ x_i,\ u(t_i\rightarrow t_f-1\ ))=E \bigg\{ L_T(x_{t_f})+\sum_{t = 0}^{t_f-1} L(t,x_t,u_t,w_t) \bigg\}
\end{equation}
and objective function:

\begin{equation}
    J(t,\ x_t)= \min_{u(t\rightarrow t_f-1)} \Bigg( E \bigg\{\sum_{t = 0}^{t_f-1} L(t,x_t,u_t, w_t) + J(t+1,\ x_{t+1}) \bigg\} \Bigg)\ , \ \forall \ x_t
\end{equation}
Again, these equations only hold for problems with a resulting end state at time \( t_f \). The discounted infinite horizon version of \autoref{equation8} is then:

\begin{align}
        C(t_i,\ x_i,\ u(t_i\rightarrow \infty)) &= \lim_{t_f \rightarrow \infty} E \bigg\{ L_T(x_{t_f})+\sum_{t = 0}^{t_f-1} L(t,x_t,u_t,w_t) \bigg\} \\[1em]
        C(t_i,\ x_i,\ u_t) &=  E \bigg\{\sum_{t = 0}^{\infty} \gamma^t L(t,x_t,u_t,w_t) \bigg\}
\end{align}
and resulting in the discounted (if \( \gamma^t < 1\)) infinite-horizon objective function:

\begin{equation}
    J_\infty(x_t)= \min_{u_t \ \in \ U_t} \Bigg( E \bigg\{\sum_{t = 0}^{\infty} \gamma^t L(t,x_t,u_t, w_t) \bigg\} \Bigg)\ , \ \forall \ x_t \label{equation12}
\end{equation}

\section{Problem statement - Energy Management Systems} \label{chapter4}
\par As shown in \autoref{table: SoA summary}, energy management systems typically utilize model predictive control techniques and, more recently, reinforcement learning. Both of them used to solve a discrete-time sequential decision making problem, as the continuous unconstrained error handling is performed by classic proportional–integral–derivative (PID) controllers (see \autoref{fig:blockdiagrams}). This section briefly discusses these approaches and their relation to the optimal control problem, i.e. \autoref{chapter3}.

\subsection{Model-predictive control}
The standard formulation of this \textit{receding horizon control} problem is:
\begin{subequations}
\begin{align}
    \label{equation13a}
    & \min_{u(t\rightarrow N-1)} \Bigg( E \bigg\{ L_{t+N}(x_{t+N})+\sum_{k = 0}^{N-1} L(t+k,x_{t+k},u_{t+k},w_{t+k}) \bigg\} \Bigg) \\[1em]
    \label{equation13b}
    s.t. \quad & x_{t+k+1}=f(t+k,\ x_{t+k},\ u_{t+k},\ w_{t+k}) && \hspace{-60pt} k \in \mathbb{N}_{0}^{N-1} &&&& \hspace{-20pt} = 0,\dots,N-1 \\
    \label{equation13c}
    & x_{t+k} \in X && \hspace{-60pt} k \in \mathbb{N}_{1}^{N} &&&& \hspace{-20pt} = 1,\dots,N \\
    \label{equation13d}
    & u_{t+k} \in U && \hspace{-60pt} k \in \mathbb{N}_{0}^{N-1} &&&& \hspace{-20pt} = 0,\dots,N-1 \\
    \label{equation13e}
    & x_0 = x(t)
\end{align}
\end{subequations}
where \(x_t \in \mathbb{R}^n \), an \textit{n}-dimensional \textit{state} vector subject to the constraint set \( X \), \(u_t \in \mathbb{R}^m \), an \textit{m}-dimensional \textit{action} or \textit{control} vector subject to the constraint set \( U \) and \(w_t\) a Wiener process, all at the \textit{t}-th step over prediction horizon \( N \) with \textit{a priori} known model \(x_{t+1} = f(t,\ x_t,\ u_t,\ w_t) \). 
\par Notice that \autoref{equation13a} is just the bounded version of the discrete time-varying stochastic optimal control problem where \( \mathbb{N}_0^{N-1} \subset \mathbb{T}_0^{t_f-1} \subset \mathbb{T}_0^{+\infty} \). This means that the solution \( \{ u_0^*,u_1^*, ..., u_{N-1}^* \} \) may be optimal over the prediction horizon  \( N \). It is always suboptimal with respect to the finite or infinite horizon of the underlying (typically \textit{continuous}) system. It is almost needless to say that this is intentional because of computational complexity considerations.

\subsection{Reinforcement learning}
\par A fully observable discrete-time \textit{Markov Decision Process} (MDP) is the standard formulation of a single-agent reinforcement learning problem and is a tuple \( \langle S,A,P_a,R_a \rangle \) where:
\begin{itemize}
    \item \( S \) represents a finite or infinite number of environment states, the \textit{state space};
    \item \( A \) represents a finite or infinite set of actions, the \textit{action space};
    \item \( P_a(s,s') = Pr(s_{t+1} = s' \ | \ s_t = s, a_t = a) \) represents a transition probability, that only depends on the current state and not on the previous ones (i.e. Markov Property), being in a certain state \( s \in S \) in time step \( t \) and choosing action \( a \in A \) will lead to being in state \( s' \in S \) in the next time step \( t+1 \);
    \item \( R_a(s,s') \) is an (expected) immediate numerical reward associated for making the transition \( s \rightarrow s' \), i.e. this is also unknown to the agent before it actually made this transition to the next state in time step \( t+1 \).
\end{itemize}
\par The objective is to find a policy \( \pi \) that maximizes an expected sum of discounted rewards, i.e. in the form of maximizing the state-\textit{value function}:
\begin{subequations}
\begin{align}
    \label{equation14a}
    \max_\pi &\Bigg(E_\pi\bigg\{ \sum_{k = 0}^{\infty} \gamma^k R_{t+k+1} \bigg\}\Bigg)\\[1em]
    \label{equation14b}
    & s.t. \quad s_t = s & t \in \mathbb{T}_{0}^{+\infty} = 0,\dots,+\infty
\end{align}
\end{subequations}
\par Note that the reward function is the negative version of the loss function \( R_{t} = -L(t, x_t, u_t, w_t) \) so that \autoref{equation14a} is equivalent to the discrete time-invariant infinite-horizon stochastic optimal control problem, i.e. the time-invariant version of \autoref{equation12}. However, no \textit{a priori} model of the system dynamics is used nor is it subject to a constraint set as is the case in model predictive control. The only knowledge required in the reinforcement learning approach is the definition of a Markov Decision Process (MDP), as presented in this section, without explicitly defining the transition probability matrix \(P_a\). The equivalence of reinforcement learning and stochastic optimal control has also been argued by \cite{Powell2019FromDecisions}, where the relationship with model-predictive control has also been discussed by \cite{Gorges2017RelationsLearning, Ernst2007ModelFrameworks}.

\section{Methodology} \label{chapter5}

\par Two multi-energy system simulation models were developed to systematically test both controllers (LMPC and RL) in an energy management context. With this purpose, \texttt{Modelica} \cite{Mattsson1998PhysicalModelica} was chosen as proposed by \cite{Graber2017FromProblems} - since it provides a general tool-chain for optimal control problems. It allows for convenient (i.e. object-oriented, elementary components, highly specialized libraries) construction of multi-physical first-principle equations  that describe the \textit{presumed} system dynamics.

\par
\begin{figure}[ht]
    \centering
    \includegraphics[scale=0.125]{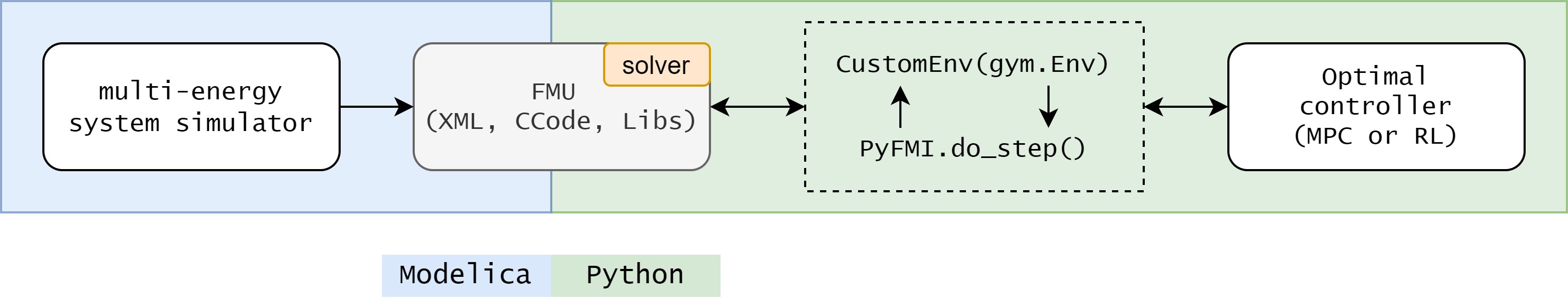}
    \caption{architecture of the tool-chain}
    \label{fig: architecture}
\end{figure}

 This \texttt{Modelica} model is then exported as a co-simulation \textit{functional mock-up unit} (FMU) and wrapped into an \texttt{OpenAI} gym \textit{environment} \cite{Brockman2016OpenAIGym} in \texttt{Python}, similar to \cite{Lukianykhin2019ModelicaGym:Models}. The overall architecture is shown in \autoref{fig: architecture}. Notice that in the co-simulation FMU, the DAE solver is wrapped as well and that the \texttt{do\_step()} method in \texttt{PyFMI} \cite{Andersson2016PyFMI:Interface} is used over \texttt{simulate()}. This requires proper initialization handling, but speeds up the simulation run time significantly. Finally, the \texttt{OpenAI} gym interface proved to be general enough to allow for the connection between any (optimal) controller and any environment.

\par In the attempt to come to a broader conclusion, both a simple and a more complex multi-energy system configurations are considered as separate case studies. These case studies are purely theoretical, however, as their main goal is to serve as challenging (optimal) control environments within the energy domain. The dimensions of the energy assets and the geographical location are therefore arbitrarily chosen. The dimension is chosen considering technical and practical limitations while maintaining all assets' relevance in the problem, which allows impactful control decisions. Germany is chosen as the location for both case studies, providing coherent data of external demand (electrical and thermal), market (natural gas and electricity prices), and weather conditions (solar and wind).

\subsection{Case study I - simple} \label{chapter5.2}
\subsubsection{Simulation model}
\par
\begin{figure}[!hb]
	\centering
	\includegraphics[scale=0.1]{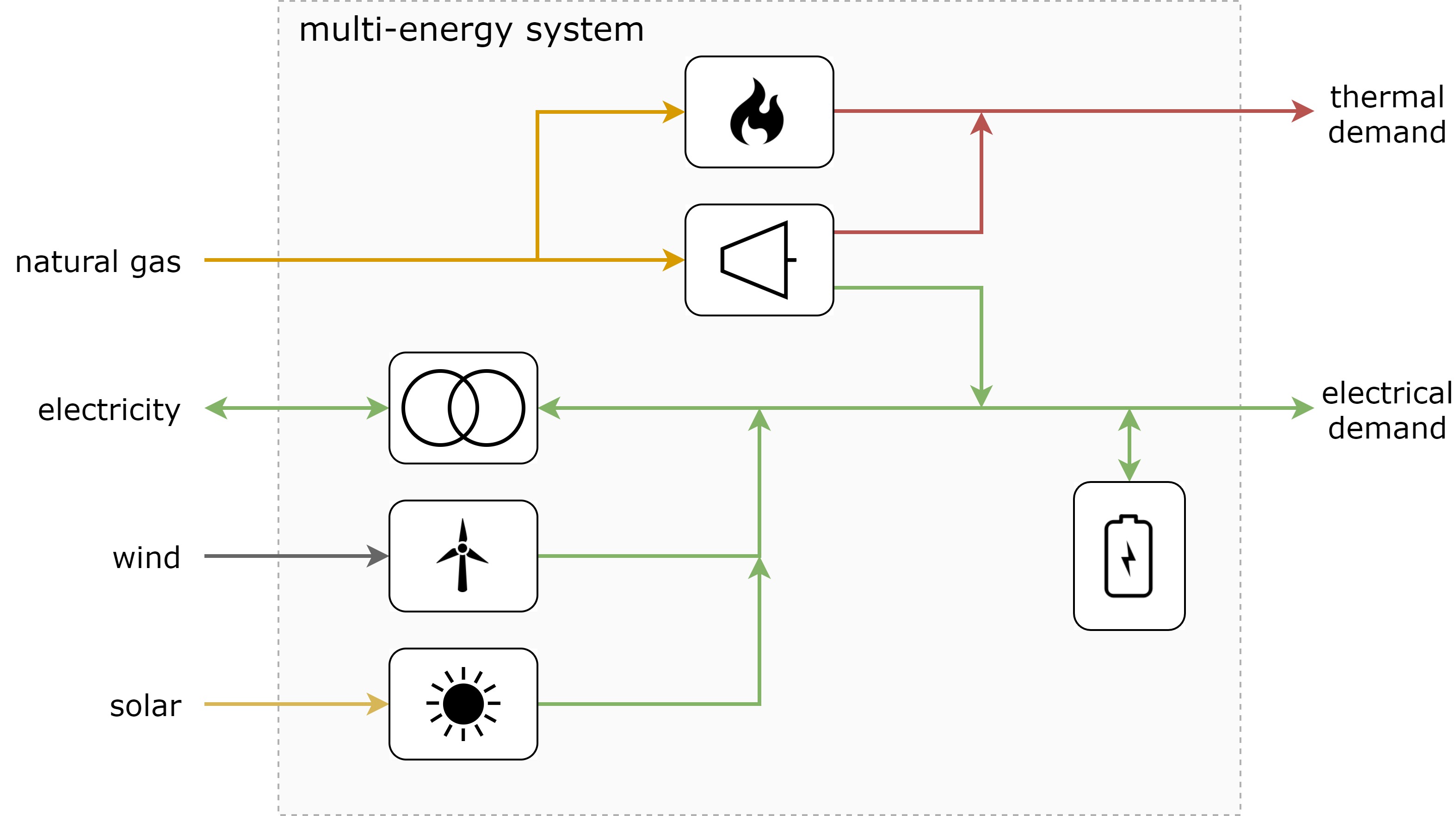}
	\caption{structure of the simple multi-energy system}
	\label{fig: case I structure}
\end{figure}

 Figure \ref{fig: case I structure} shows a simplified and visual representation that describes the interaction between the 2430 differential-algebraic equations that rule the dynamics of the simple multi-energy system. The model is based on an example provided by the open-source Modelica library \texttt{TransiENT} \cite{TransiENT}. However, several modifications have been applied to the original example, aiming for a more complete model e.g. added electricity market information and the introduction of storage models. The system is comprised of white-box models of a wind turbine, a photovoltaic (PV) generation plant, a natural gas boiler, a combined heat and power (CHP) unit and a battery energy storage system (BESS). 
\par Each energy asset in the system is defined by a subset of the DAEs mentioned before. However, the description of these equations is out of the scope of this paper. As a brief explanation, the dynamics of the system include a descriptive model of an extraction CHP that allows the plant to operate on a feasible PQ surface (different combinations of electric and heat power outputs). Each of these combinations is associated with an electrical efficiency, thermal efficiency, and gas input required. Furthermore, the battery energy storage system includes state-of-charge dependent maximum rates of charge/discharge. Both renewable electricity generation plants are non-controllable in this model. Finally, the inputs/outputs, nominal capacities and safe operational limits are described in \autoref{tab: simple mes dimensions}. 
\begin{table}[!hb]
    \centering
    \begin{tabular}{c|c|c|c|c|c}
    \rowcolor[HTML]{efefef} 
    \textbf{Energy asset} & \textbf{Input} & \textbf{Output} & \textbf{P\textsubscript{nom}} & \textbf{P\textsubscript{min}} & \textbf{E\textsubscript{nom}} \\
    \hline
    wind turbine & wind & elec & 5.0 MW\textsubscript{e} & 0 \% & \\
    solar PV & solar & elec & 3.0 MW\textsubscript{e} & 0 \% & \\
    boiler & CH\textsubscript{4} & heat & 8.0 MW\textsubscript{th} & 0 \% & \\
    CHP & CH\textsubscript{4} & heat & 6.0 MW\textsubscript{th} & 25 \% & \\
     & CH\textsubscript{4} & elec & 6.0 MW\textsubscript{e} & 25 \% & \\
    BESS & elec & elec & +2.5 MW\textsubscript{e} & -2.5 MW\textsubscript{e} & 10.0 MWh 
    \end{tabular}
    \caption{dimensions of the simple multi-energy system}
    \label{tab: simple mes dimensions}
\end{table}

\subsubsection{Model-predictive controller}
\par In this case, \texttt{ficus} \cite{ficus} was selected to model and solve the receding optimal control problem, formulated as a mixed-integer linear program using CPLEX as the solver. Even though the modelling framework differs from the one used in the following case, the optimisation problem is still based on the formulation detailed in \autoref{equation15a}-\ref{equation15e}.
\begin{subequations}
\begin{align}
    \label{equation15a}
    \min_{u(a,k\rightarrow A,N-1)} & \bigg( \sum_{a = 0, k = 0}^{A,N-1} L_{cost}(t+k,\ x_{a,t+k},\ u_{a,t+k}) \bigg) \\[1em]    
    \label{equation15b}
    s.t. \quad & x_{a,t+k+1}=f(t+k,\ x_{a,t+k},\ u_{a,t+k}) && a \in \mathbb{A}_{1}^{A}, k \in \mathbb{N}_{0}^{N-1} \\
    \label{equation15c}
    & x_{a,t+k} \in X && a \in \mathbb{A}_{1}^{A}, k \in \mathbb{N}_{1}^{N} \\
    \label{equation15d}
    & u_{a,t+k} \in U && a \in \mathbb{A}_{1}^{A}, k \in \mathbb{N}_{0}^{N-1} \\
    \label{equation15e}
    & x_{a,0} = x_a(t)
\end{align}
\end{subequations}
\par Where \( L_{cost} \) is the operational energy cost loss function subject to the constraint sets \(X\) and \(U\), at the \textit{t}-th step over the prediction horizon \( N \). This formulation has the \textit{a priori} known discrete \textit{deterministic} model \( x_{a,t+k+1} = f(t+k,\ x_{a,t+k},\ u_{a,t+k}) \). The multi-dimensional decision variable \(u\) contains the set-point for a number \(A\) of assets \(a\) from set \(\mathbb{A}\) for each time-step \(t\) in the prediction horizon \(\mathbb{N}_{0}^{N-1}\), which results in a mixed-integer linear optimisation program (MILP).
\par It is important to highlight that the model from \texttt{ficus} represents some of the dynamics of the actual system in a linearized and simplified way, thus, allowing the use of a MILP solver. Consequently, the solution obtained slightly deviates from the optimum. Nevertheless, this is still considered an adequate solution since most of the mathematical models in different studies are subject to this linearization to reduce computation time and ensure the existence of a solution. Therefore the use of a \textit{linear} MPC as a benchmark is justified, though better-performing nonlinear MPCs exist as well \cite{Jorissen2021NMPC, CupeiroFigueroaIago2020AMfL}.
\par The LMPC is operated in a receding horizon control fashion as depicted in \autoref{algo: mpc}. The prediction horizon is set to be 72 hours (related to the number of steps in the prediction horizon \(N\)) while a 24 hour period is taken as the control horizon (where \(C\) is the number of steps in the control horizon), using a time-step interval \(T_s = 15\) minutes. Therefore, an optimization problem is solved by taking the forecasts for the next three days and selecting the values of the decision variables that optimize the objective function. The solution is passed to the simulation environment for a time interval equal to the control horizon. Once the control horizon ends, the state variables are updated directly by measuring the state at the end of this horizon. This process is repeated until the simulation period finishes (365 simulation days). A control horizon of 15 minutes was also tested (i.e. so that \(C = T_s\)), yet no significant difference in performance was found in these case studies.
\begin{algorithm}
\DontPrintSemicolon
\SetAlgoLined
 \nl initialization\;
 \nl \For{\( t = 0, C, 2C, \ldots, +\infty \)}{
 \nl Measure the initial state \(x_{t}\) \;
 \nl Predict the (non action-dependent) states \(\{\hat{x}_{t},\hat{x}_{t+1},\ldots,\hat{x}_{t+N-1}\}\) over horizon \(N\) \;
 \nl  Solve \autoref{equation15a}-\ref{equation15e} to compute the optimal sequence of control actions \(\{u_{t}^*,u_{t+1}^*,\ldots,u_{t+N-1}^*\}\)\;
 \nl  Select the first C elements \(\{u_{t}^*,u_{t+1}^*,\ldots,u_{t+C-1}^*\}\) \;
 \nl \For{\( k = t, t+1, \ldots, t+C-1 \)}{
 \nl  Apply the control action \(u_{k}^*\) to the system \;
 }
 }
 \caption{MPC}
 \label{algo: mpc}
\end{algorithm}

\begin{wraptable}{r}{6.7cm}
\vspace{-15pt}
    \centering
    \begin{tabular}{l|c|c}
    \rowcolor[HTML]{efefef} 
    \textbf{Predictions} & \textbf{R2-score} & \textbf{MAPE} \\
    \hline
    Electrical demand & 95\% & 4\% \\
    Thermal demand & 94\% & 9\% \\
    Day-ahead price & 86\% & 6\% \\
    Solar irradiance & 69\% & 37\%\footnotemark[2] \\
    Wind speed & 78\% & 25\%\footnotemark[2]
    \end{tabular}
    \caption{imperfect LMPC prediction metrics for the simple MES configuration}
    \label{tab: prediction errors simple}
\vspace{-15pt}
\end{wraptable}
\par The results are presented for two different scenarios. On the one hand, the receding horizon optimisation is solved with perfect knowledge of the future energy prices, renewable generation, and thermal and electrical demands. This sets the best performance that the \textit{ficus} model can achieve in this custom simulation model. On the other hand, Gaussian noise (the forecasting error is assumed to have a normal distribution) is added to the states in the prediction horizon, to mimic the realistic operation of an MPC. This alteration results in an imperfect forecast of future states, which is commonly found in model predictive controllers. More precisely, the state variables that are mixed with the Gaussian noise are the future electricity prices, wind speeds, solar irradiation and both thermal and electrical loads. These variables are modified according to the metrics presented in \autoref{tab: prediction errors simple}, which is assumed to resemble state-of-the-art 3-day-ahead forecasting performance with quarterly intervals. 

\subsubsection{Reinforcement learning agent}
\par For the simple multi-energy system, the \texttt{PPO2} implementation of OpenAI is used. Here, the fully observable discrete-time MDP is formulated as the tuple  \( \langle S,A,P_a,R_a \rangle \):
\footnotetext[2]{The exclusion of zeros causes the high values}
\begin{subequations}
\begin{align}
    S^t = ( E_{th}^t,\ E_{el}^t,\ P_{wind}^t,\ P_{solar}^t,\ C_{e}^t,\ X_{el}^t  )& & S^t \in S\\[1em]
    A_{chp, p}^t = (0,\ A_{chp, p}^{min} \rightarrow A_{chp, p}^{max})& & A_{chp, p}^t \in A \\
    A_{chp, q}^t = (0,\ A_{chp, q}^{min} \rightarrow A_{chp, q}^{max})& & A_{chp, q}^t \in A \\   
    A_{bess}^t = (A_{bess}^{min} \rightarrow A_{bess}^{max})& & A_{bess}^t \in A \\[1em]
    \label{rew_c_II}
    R_a = - (a \times L_{cost}^t + b \times L_{comfort}^t)
\end{align}
\end{subequations}

where \( E_{th}^t \) is the thermal demand, \( E_{el}^t \) the electrical demand, \( P_{wind}^t \) the electrical wind in-feed, \( P_{solar}^t \) the electrical solar in-feed, \( C_{e}^t \) the total energy cost and \(X_{el}^t\) the electrical price signal (i.e. day-ahead spot price) all at the \textit{t}-th step which makes up the state-space \( S \). The continuous action-space \( A \), is comprised of the combined heat and power electric \( A_{chp,p}^t \) and thermal \( A_{chp,q}^t \) control set-points, alongside with the battery energy storage system electric power, all within the power rate limits shown in \autoref{tab: simple mes dimensions}. 
\par It is important to mention that differently to the following case study, the gas boiler is not directly controllable since its set-point is determined by a separate controller that automatically chooses as set-point the remaining thermal demand after the CHP heat production. This decision was based on the avoidance of redundancy in the controller since the heat not supplied by the CHP must be supplied by the boiler to fulfil the thermal demand. Hence, the action-space could be kept in a lower dimension space, improving the learning process of the RL agents. 
\par Finally, the reward function \( R_a \) is the negative loss in energy costs \( L_{cost}^t \) and loss in comfort \( L_{comfort}^t \) both at the \textit{t}-th time-step with multi-objective scaling constants \( a \) and \( b \). The loss in comfort is defined as \( | E_{th}^t - Q^t | \), where \( Q^t \) is the thermal energy production. The electrical demand and natural gas consumption can always be fulfilled by (buying from) their respective \textit{infinitely} large main grid connection, i.e. within the \texttt{Modelica} simulation model, it is assumed that the grid connections are sufficiently large. 

\subsection{Case study II - complex} \label{chapter5.1}
\subsubsection{Simulation model} \label{chapter5.1.1}

\par The considered complex multi-energy system has the following structure:
\begin{figure}[H]
	\centering
	\includegraphics[scale=0.1]{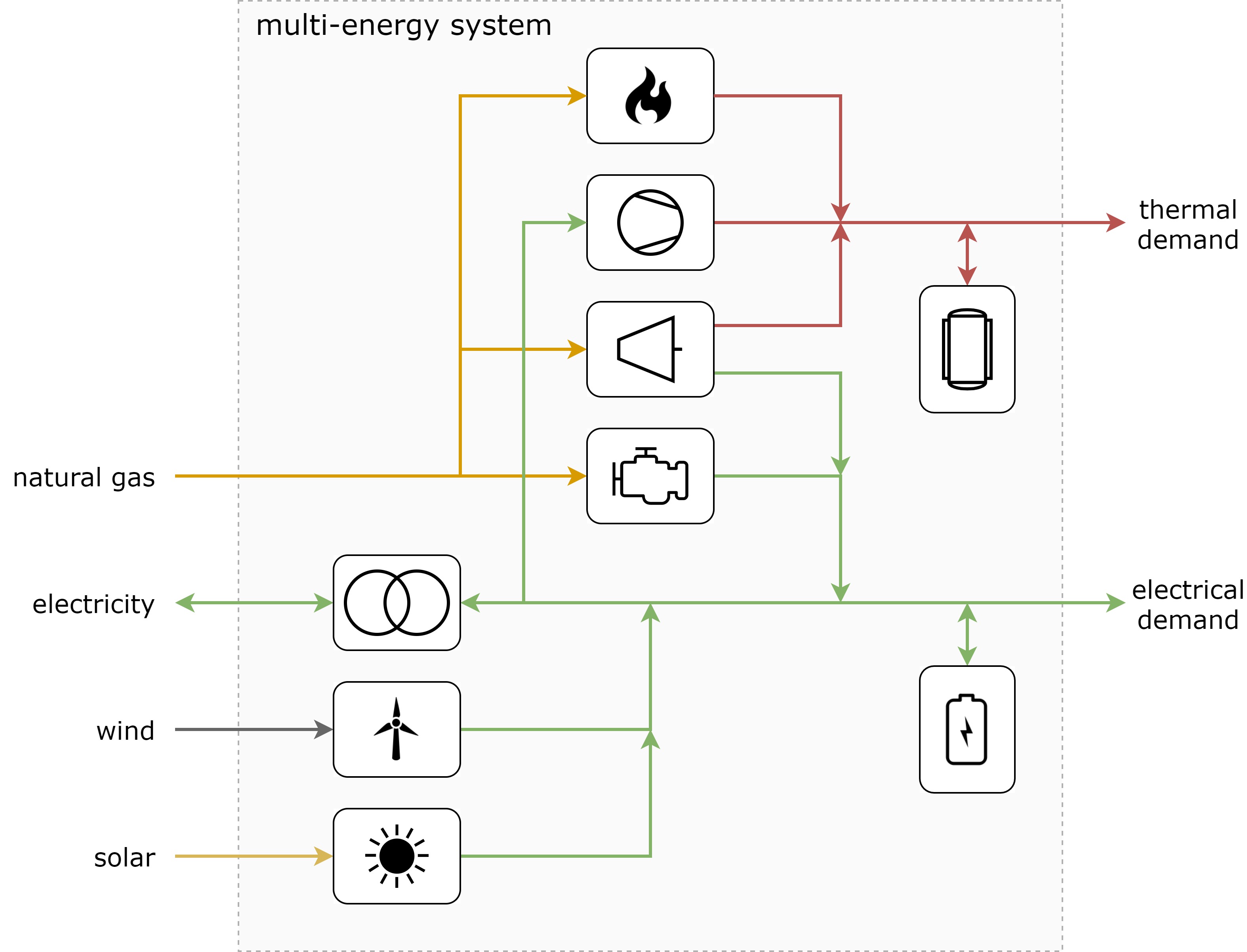}
	\caption{structure of the complex multi-energy system}
	\label{fig: mes structure}
\end{figure}

\par It includes (from left to right, from top to bottom): an electric transformer, a wind turbine, a photovoltaic (PV) installation, a natural gas boiler, a heat pump, a combined heat and power (CHP) unit, a gas turbine (back-up) generator, a thermal energy storage system (TESS) and a battery energy storage system (BESS). Notice that there are multiple assets that link the energy carriers together, requiring an integrated control strategy - hence the reason for this presented system configuration. The dimensions of the considered multi-energy system are presented in \autoref{tab: mes dimensions}.
\begin{table}[!ht]
    \centering
    \begin{tabular}{c|c|c|c|c|c}
    \rowcolor[HTML]{efefef} 
    \textbf{Energy asset} & \textbf{Input} & \textbf{Output} & \textbf{P\textsubscript{nom}} & \textbf{P\textsubscript{min}} & \textbf{E\textsubscript{nom}} \\
    \hline
    transformer & elec & elec & \(+\infty\) & \(-\infty\) & \\
    wind turbine & wind & elec & 0.8 MW\textsubscript{e} & 1.5 \% & \\
    solar PV & solar & elec & 1.0 MW\textsubscript{e} & 0 \% & \\
    boiler & CH\textsubscript{4} & heat & 2.0 MW\textsubscript{th} & 10 \% & \\
    heat pump & elec & heat & 1.0 MW\textsubscript{th} & 25 \% & \\
    CHP & CH\textsubscript{4} & heat & 1.0 MW\textsubscript{th} & 50 \% & \\
     & CH\textsubscript{4} & elec & 0.8 MW\textsubscript{e} & 50 \% & \\
    genset & CH\textsubscript{4} & elec & 0.5 MW\textsubscript{e} & 50 \% & \\
    TESS & heat & heat & +0.5 MW\textsubscript{th} & -0.5 MW\textsubscript{th} & 3.5 MWh \\
    BESS & elec & elec & +0.5 MW\textsubscript{e} & -0.5 MW\textsubscript{e} & 2.0 MWh 
    \end{tabular}
    \caption{dimensions of the multi-energy system}
    \label{tab: mes dimensions}
\end{table}

\par Do note that this table is a simplified, yet convenient, representation of the underlying system characteristics and its scale. For example, the minimum wind turbine power is a result of a minimum cut-off wind speed and an efficiency curve, and the thermal energy storage system is a given volume of water operating between certain temperature ranges of a hydraulic heating system. In this case study, assumed to be in Dresden (Germany), the aggregated thermal and electrical hourly demand of 100 households is used and is linearly interpolated between time-steps. Day-ahead electricity prices are downloaded from the platform of the European Network of Transmission System Operators (entso) \cite{ENTSO-E2018ENTSO-EPlatform} and weather data from \cite{EnergyPlus2016WeatherEnergyPlus}. However, it is beyond the scope of this paper to (mathematically) describe this simulation model in detail, yet it is a system of 2,548 differential-algebraic equations (DAE) and thus with an equal amount of variables.

\subsubsection{Model-predictive controller}
\par Many open-source energy system optimisation tools exist \cite{Groissbock2019AreUse}. In contrast to case I, the multi-scale energy systems modelling framework \texttt{calliope} \cite{Pfenninger2018Calliope:Framework} was chosen, which provides convenient classes to use out-of-the-box (e.g. receding control horizon). The objective function of the model-predictive controller is the same as for the previous case, and is defined in \autoref{equation15a}-\ref{equation15e}.
\par The most restrictive element of this model is the need to balance the power of the respective carriers on every time step. Thus, the power that technologies produce, storage unit power output and power imports must equal the sum of consumed power, power placed in storage, power exports and the systems power demands. Hence, the energy management objective is to minimize energy costs while fulfilling this comfort constraint. These (linear) equations and all subsets, parameters, and variables can be found in the mathematical documentation of \texttt{calliope} \cite{MathematicalDocumentation}. 

\begin{wraptable}{r}{6.7cm}
\vspace{-15pt}
    \centering
    \begin{tabular}{l|c|c}
    \rowcolor[HTML]{efefef} 
    \textbf{Predictions} & \textbf{R2-score} & \textbf{MAPE} \\
    \hline
    Electrical demand & 95\% & 12\% \\
    Thermal demand & 95\% & 8\% \\
    Day-ahead price & 85\% & 8\% \\
    Solar irradiance & 70\% & 37\%\footnotemark[2] \\
    Wind speed & 70\% & 36\%\footnotemark[2]
    \end{tabular}
    \caption{imperfect LMPC prediction metrics for the complex MES configuration}
    \label{tab: prediction errors complex}
\vspace{-15pt}
\end{wraptable}

\par Also in this case study, a prediction horizon of 72 hours and a control horizon of 24 hours were chosen with a discretization of 15 minutes, both with perfect and imperfect foresight. The pseudo-code is given in \autoref{algo: mpc}. The \textit{assumed} metrics of the imperfect predictions, independent of any modelling bias (e.g. presumed efficiency curves), are presented in \autoref{tab: prediction errors complex}.

\subsubsection{Reinforcement learning agent}
The fully observable discrete-time MDP is formulated as the tuple \( \langle S,A,P_a,R_a \rangle \) so that:
\begin{subequations}
\begin{align}
    S^t = ( E_{th}^t,\ E_{el}^t,\ P_{wind}^t,\ P_{solar}^t,\ C_{e}^t,\ X_{el}^t  )& & S^t \in S\\[1em]
    A_{boil}^t = (0,\ A_{boil}^{min} \xrightarrow{\tau} A_{boil}^{max})& & A_{boil}^t \in A \\
    A_{hp}^t = (0,\ A_{hp}^{min} \xrightarrow{\tau} A_{hp}^{max})& & A_{hp}^t \in A \\
    A_{chp}^t = (0,\ A_{chp}^{min} \xrightarrow{\tau} A_{chp}^{max})& & A_{chp}^t \in A \\
    A_{gen}^t = (0,\ A_{gen}^{min} \xrightarrow{\tau} A_{gen}^{max})& & A_{gen}^t \in A \\
    A_{tess}^t = (A_{tess}^{min} \xrightarrow{\tau} A_{tess}^{max})& & A_{tess}^t \in A \\
    A_{bess}^t = (A_{bess}^{min} \xrightarrow{\tau} A_{bess}^{max})& & A_{bess}^t \in A \\[1em]
    R_a = - (a \times L_{cost}^t + b \times L_{comfort}^t)&
\end{align}
\end{subequations}

\footnotetext[2]{The exclusion of zeros causes the high values}

where the state-space \( S \) and the reward function \( R_a \) are the same as in the previous case study. The action-space \( A \), with discretization constant \( \tau \), is confined of the control set-points from, \( A_{boil}^t \) the natural gas boiler, \( A_{hp}^t \) the heat pump, \( A_{chp}^t \) the combined heat and power unit, \( A_{gen}^t \) the natural gas gen-set, \( A_{tess}^t \) the thermal energy storage system and \( A_{bess}^t \) the battery storage system all between a minimum and maximum power rate as shown in \autoref{tab: mes dimensions}. 
\par Here, proximal policy optimisation (PPO) \cite{Schulman2017ProximalAlgorithms} and twin delayed deep deterministic policy gradient (TD3) \cite{Fujimoto2018AddressingMethods} agents were chosen as they are considered one of the state-of-the-art model-free reinforcement learning algorithms, on- and off-policy respectively. Specifically, the \texttt{PPO2}, with discrete action-space, and \texttt{TD3}, with continuous action-space, implementations of the \texttt{stable baselines} \cite{stable-baselines} are used. The pseudo-code is given in \autoref{algo: ppo} and \autoref{algo: td3} in \hyperref[Appendix D]{Appendix~\ref*{Appendix D}} and the hyper-parameter study is described in \hyperref[Appendix B]{Appendix~\ref*{Appendix B}}. All appendices are located in the supplementary material.

\section{Results and discussion} \label{chapter7}
\par This section presents the performance, in terms of energy cost minimization, subject to comfort fulfilment of the perfect and realistic foresight (\autoref{tab: prediction errors simple}, \autoref{tab: prediction errors complex}) linear model-predictive controller, the reinforcement learning agents (\autoref{algo: ppo} and \autoref{algo: td3}) and random agents (as minimal performance baseline), presented relative towards the perfect foresight LMPC. The optimal control policies are evaluated over a year-long simulation while participating in a day-ahead electricity market.

\par The results observed from the simulations show the objective value relative to the simplified LMPC with perfect foresight and a finite receding optimisation horizon. These values are a metric for the \textit{multi-task} energy management problem of minimizing the energy cost while fulfilling the demands and are presented in \autoref{tab: case results}. Detailed visualisations of the simulations' results (i.e. time series plots of the control policies) can be found in \hyperref[Appendix C]{Appendix~\ref*{Appendix C}}. 

\begin{table}[!htbp]
    \centering
    \begin{tabular}{l|c|c}
    \rowcolor[HTML]{efefef} 
    \textbf{Optimal controller} & \textbf{Case study I} & \textbf{Case study II}\\
    \hline
    LMPC - perfect predictions & 100\% & 100\% \\
    LMPC - realistic predictions&  98.0\% & 89.9\% \\
    PPO agent & 101.3\% & 88.1\% \\ 
    TD3 agent & 101.5\% & 94.6\% \\
    Random agent & 82.4\%\footnotemark[4] & 12.8\%    \end{tabular}
    \caption{simulation results of the case studies.}
    \label{tab: case results}
\end{table}

\footnotetext[4]{Random agent + boiler controller}

\par These results (\autoref{tab: case results}) show that a significant amount of performance is lost due to inaccurate prediction \textit{and} modelling errors in the linear model predictive controller (LMPC), i.e. 2\% in case study I and 10\% in case study II. This occurs since even the perfect foresight LMPC is still a simplified or approximated (only linear equations) representation of the true underlying system dynamics. Furthermore, these results show that the reinforcement learning (RL) agents can even outperform the perfect foresight LMPC (in the simple MES configuration). This, as they make no \textit{a priori} assumptions on the underlying system dynamics nor they require a state forecast as input and, thus, learn a better representation of those systems' dynamics given a sufficient training budget (i.e., a finite amount of interaction with the environment). 
\par The economic viability of the energy management system even increases as no case-specific, \textit{a prior} models need to be constructed and maintained (yet training effort is needed). Nevertheless, the RL agents by themselves rely on multiple function approximations (e.g. using artificial neural networks) always resulting in a sub-optimal solution (arguably the practically achievable optimal). In particular, the RL agents showed difficulties in operating both the thermal and electrical storage system as this involves momentarily a penalty (i.e. increased consumption when charging) to achieve a mid-term reward (discharging when prices are higher). Future work will assess whether this is caused by the selected hyper-parameters (\hyperref[Appendix B]{Appendix~\ref*{Appendix B}}, gamma is already considered in this paper), the surrogate objective implemented in PPO, the use of PPO as the reinforcement learning algorithm or the use of reinforcement learning itself.
\par Moreover notice that the random agent in case study I shows a high relative performance as compared with case study II. This is not only because of the simpler multi-energy system configuration but also due to the separate controller in the boiler, causing the thermal discomfort to be \textit{close to zero} even when the set-points in the rest of the system are random. This has a considerable impact on the thermal discomfort term in \autoref{rew_c_II}, minimizing its effect and causing a relatively high performance.
\par However, the results of the \textit{unconstrained} (apart from the finite action space) RL agents are achieved by a large number of unsafe interactions with its environment, resulting in a training cost (that does not exist in the LMPC, yet has an initial and ongoing modelling cost). This is observed for both tasks: minimizing the energy cost (i.e., expending more than needed) and fulfilling the demands (i.e. the thermal system is not grid-connected, resulting in thermal discomfort).

\begin{figure}[H]
    \centering
    \includegraphics[width=0.90\textwidth, height=0.4\textheight, trim={1cm, 1cm, 1cm, 1cm}, clip]{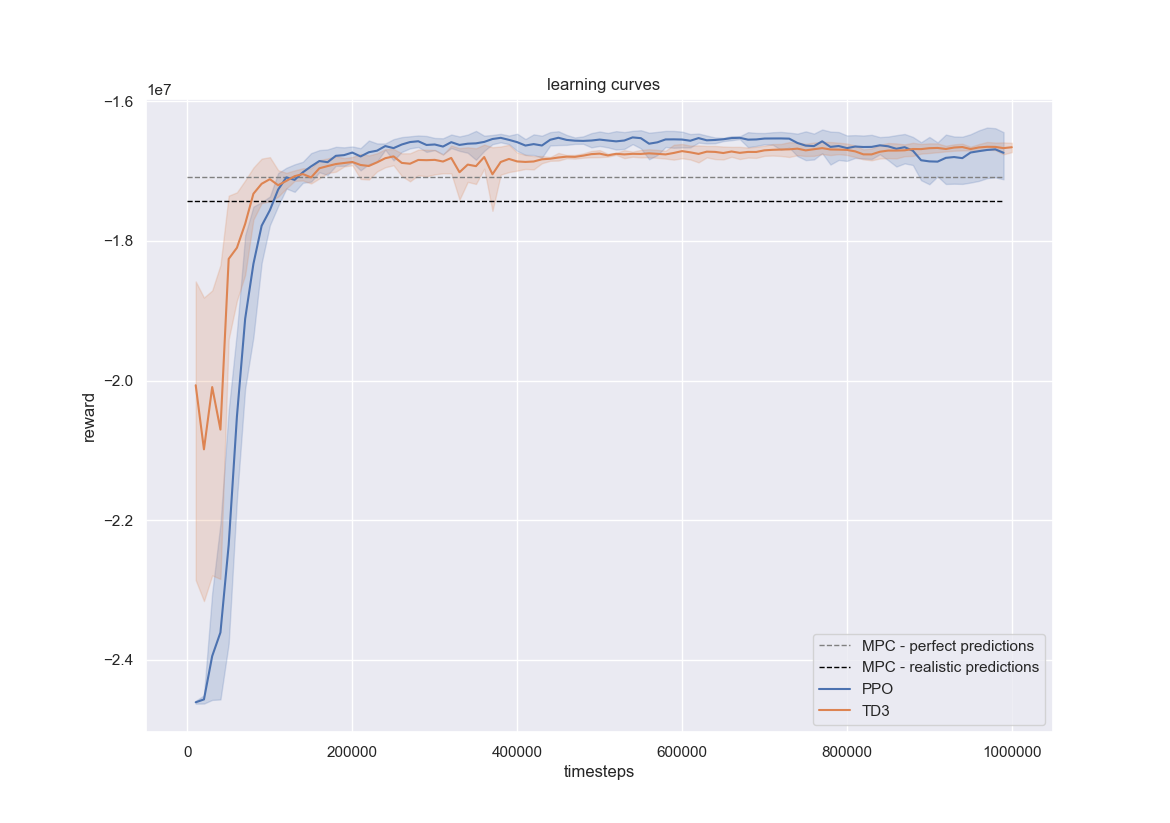}
    \caption{Case study I: learning curves}
    \label{fig: learning curve case 2}
\end{figure}

\par The learning curves of the RL agents in case study I are presented in \autoref{fig: learning curve case 2}. It shows stabilization in the reward for both agents after \(\sim5.70\) years (200,000 training time-steps). However, the multiple agents trained using PPO start to deviate, which is visible at approximately 650,000 time-steps. Opposed to that, TD3 shows a high variation during the starting steps of the training but stabilizes in the mid to long term, since the deviation between agents tends to decrease as the number of training time steps increases. Overall, both algorithms find a policy that outperforms the perfect foresight LMPC after 125,000 training steps which is equivalent to 3.56 years worth of interaction with the environment and \(\sim80,000\) and \(\sim105,000\) time-steps to outperform the realistic LMPC (2.28 and 3 years respectively). This long training time could be impractical in a real setup, which suggests that further research is needed to overcome this challenge. 
\par The time in which these agents reached 70\% of their best performance were 2.04 years for PPO and 2.17 years for TD3. Similarly, they achieved 90\% in 2.94 and 3.57 years, respectively.

\begin{figure}[H]
    \centering
    \includegraphics[width=0.90\textwidth, height=0.4\textheight, trim={1cm, 1cm, 1cm, 1cm}, clip]{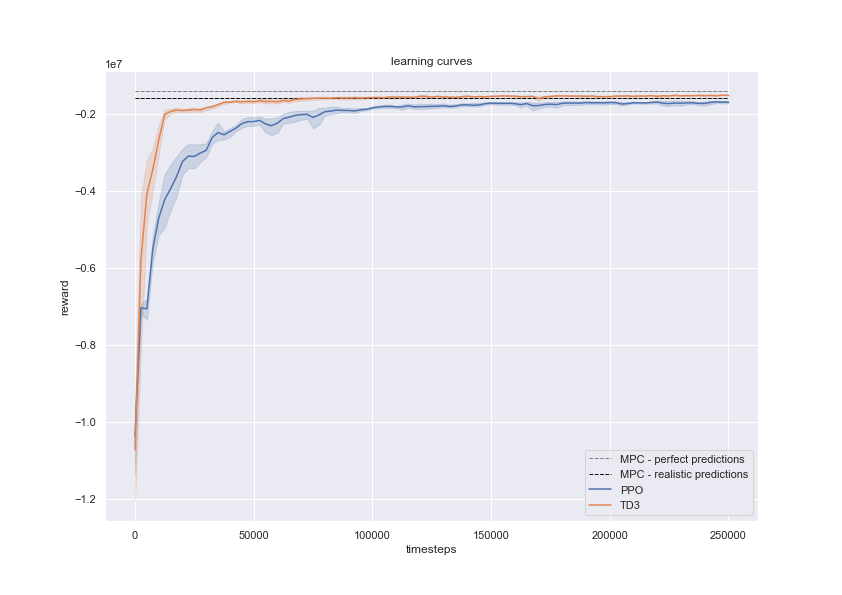}
    \caption{Case study II: learning curves}
    \label{fig: learning curve case 1}
\end{figure}

\par In \autoref{fig: learning curve case 1} the learning curves of the RL agents in case study II are presented. We observe a steep initial learning rate, a low variance, and a stable performance with an increasing number of interactions with the environment. The TD3 agent reaches \(\sim70\%\) of its maximum performance after 3,000 time steps or \(\sim4.5\) weeks of data (using quarterly interactions), \(\sim90\%\) after 10,000 time steps or \(\sim14.9\) weeks and \(\sim95\%\) after 12,500 time steps or \(\sim18.6\) weeks .The realistic LMPC is outperformed after \(\sim2\) years. It is expected that a model-based RL agent would have an improved learning speed (due to the additional planning step).



%

\section{Conclusions and future work} \label{chapter8}
\par This paper benchmarked a state-of-the-art on- and off-policy multi-objective reinforcement learning (RL) algorithm against a linear model predictive controller (LMPC) in dynamically simulated, day-ahead participating, multi-energy systems.
\par We conclude that RL can be an adequate optimal control technique for multi-energy management systems that outperforms \textit{practically} achievable LMPC, given a sufficient training budget (i.e. time and memory). However, while avoiding the initial and ongoing modelling cost that exists with the MPC technique, the superior RL performance is achieved by a large number of unsafe interactions with its environment, resulting in a training cost. 
\par Additionally, the performance greatly depends on the selection of adequate \textit{a priori} unknown hyper-parameters, yet this can be mitigated by performing offline pre-studies. Furthermore, the hyper-parameters were found to be interchangeable across multi-energy systems configurations. This statement is based on the fact that the hyper-parameter study results obtained in case II were also used in case I, achieving an agent that outperformed the benchmark. It indicates that they are task-specific and not completely environment-specific, however, further research on this statement is needed. Therefore, we propose the following future work:

\begin{itemize}
    \item Achieving asymptotic stability (i.e., safely recover from exploratory actions) that can be guaranteed by first estimating a safe region of attraction for an initial control policy using Lyapunov stability theory (i.e. comparison function). The state-space is then explored, never leaving the safe region with high probability. This by updating any statistical model (e.g., Gaussian Process) of the dynamics and safely improving the control policy.
    \item Utilizing shielding principles to enforce strict safety constraints expressed as temporal logic and to introduce domain knowledge to reduce the training budget.
    \item Including a planning step in the reinforcement learning agents to further reduce the training budget and to roll out a fixed sequence of (robust\footnotemark) control actions (e.g. using a Dyna-algorithm or native model-based RL agents) for day-ahead \textit{planning} purposes.
    \item Extensions towards demand-side management (DSM) and control of industrial production processes, including complexities in energy markets from a production planning perspective.
    \item Root-cause analysis on the limited use of the BESS in both case studies, including a review of the state-space, reward function and hyper-parameters (e.g. discount rate).
    \item Robustness of the energy management systems under faulty measurements/observations and sensitivity of its hyper-parameters across multiple system configurations.
\end{itemize}
\footnotetext{The transition probability matrix can then also be used to generate a \textit{robust} planning rather than a pure \textit{most-likelihood} planning}

\section{Acknowledgement}
This work has been supported in part by ABB n.v. and Flemish Agency for Innovation and Entrepreneurship (VLAIO) grant HBC.2019.2613 and grant HBC.2018.0529 . We also acknowledge Flanders Make for the support to the MOBI research group.

\section*{CRediT authorship contribution statement}
\par \textbf{Glenn Ceusters}: Conceptualization, Methodology, Software, Validation, Formal analysis, Resources, Data curation, Writing - original draft, Visualization, Funding acquisition; \textbf{Román Cantú Rodríguez}: Software, Formal analysis, Resources, Data curation, Writing - original draft, Visualization; \textbf{Alberte Bouso García}: Formal analysis, Data curation, Visualization; \textbf{Rüdiger Franke}: Supervision; \textbf{Geert Deconinck}: Writing - review and editing, Supervision; \textbf{Lieve Helsen}: Writing - review and editing; \textbf{Ann Nowé}: Writing - review and editing, Supervision; \textbf{Maarten Messagie}: Supervision; \textbf{Luis Ramirez Camargo}: Conceptualization, Writing - review and editing, Supervision.

\appendix
\section{Definitions}
\label{Appendix A}
\subsection{Coefficient of determination (R2-score)}

If \( \hat{y}_i \) is the predicted value of the \(i\)-th sample and \(y_i\) is the corresponding ground-truth value for total \(n\) samples, the estimated \(R^2\) is defined as:

\begin{equation}
    R^2(y, \hat{y}) = 1 - \frac{\sum_{i=1}^{n} (y_i - \hat{y}_i)^2}{\sum_{i=1}^{n} (y_i - \bar{y})^2}
\end{equation} \\
where \( \bar{y} = \frac{1}{n} \sum_{i=1}^{n} y_i \) and \( \sum_{i=1}^{n} (y_i - \hat{y}_i)^2 = \sum_{i=1}^{n} \epsilon_i^2 \) .

\subsection{Mean absolute percentage error (MAPE)}

If \( \hat{y}_i \) is the predicted value of the \(i\)-th sample and \(y_i\) is the corresponding ground truth value, then the mean absolute percentage error (MAPE) estimated over \( n_{\text{samples}} \) is defined as:

\begin{equation}
    \text{MAPE}(y, \hat{y}) = \frac{1}{n_{\text{samples}}} \sum_{i=0}^{n_{\text{samples}}-1} \frac{{}\left| y_i - \hat{y}_i \right|}{max(\epsilon, \left| y_i \right|)}
\end{equation} \\
where \( \epsilon \) is an arbitrary small yet strictly positive number to avoid undefined results when y is zero.
\newpage

\section{Hyper-parameter optimisation}
\label{Appendix B}
\par As it is known, the performance of Machine Learning techniques closely depends on a set of hyper-parameters that are selected before the learning process. These hyper-parameters influence the multi-step optimisation process that aims to develop a policy that successfully optimizes the defined objective function. Hyper-parameter tuning can be performed manually by running multiple cycles of training and evaluation (trials) using different combinations of feasible hyper-parameters \cite{Florea2019}. Note that each trial involves the training process of the policy, making this cycle highly computationally and time-intensive. Consequently, automatic tuning of these parameters is desired with the goal of reducing the total amount of training phases before finding satisfactory hyper-parameters. 
\par Automated hyper-parameter optimisation has the advantage of reducing both the human effort and human error while improving the performance of the machine learning algorithm. Additionally, the reproducibility and more importantly, the fairness of comparison is enhanced in research involving machine learning techniques \cite{Feurer2019}. 
\par There are several hyper-parameter search techniques developed in the literature, among them: Grid search, Random Search, Sequential Model-Based Global optimisation, Nelder-Mead technique, simulated annealing, evolutionary algorithms, genetic algorithms, particle swarm optimisation, Bayesian methods, etc. \cite{Florea2019, Feurer2019}. Consequently, there are software tools aimed to undergo optimal hyper-parameter search through some of these techniques such as Hyperopt, Spearmint, SMAC, Autotune, Vizier, Optunity, Google HyperTune, OptiML, SigOpt \cite{Florea2019, Akiba2019}.

\par In the case of the work developed in this text, the reinforcement learning agents in cases I and II require the specification of a set of hyper-parameters that are used in the PPO and TD3 algorithms. Depending on the values assigned to these, the simulations resulted in a diversity of learning curves. Not only their learning speeds differed but also their best performance achieved (i.e., their final policy and objective function values). Therefore, the presented results are based on RL agents trained with hyper-parameters found after running a hyper-parameter optimisation study. In this case, the python module Optuna \cite{Akiba2019} is selected to complete this optimisation task. Optuna is an optimisation software designed specifically with the objective of running hyper-parameter optimisation studies for machine learning algorithms. It allows the user to define a custom search space over which the algorithm finds the set of parameters that optimizes a custom objective function (also set by the user). For both case studies, a Tree-structured Parzen Estimator (TPE) is used as the hyper-parameter optimisation search algorithm which is described in \cite{BergstraJames2011AfHO}.

\begin{table}[!hb]
    \centering
    \begin{tabular}{l|c|c}
    \rowcolor[HTML]{efefef} 
    \textbf{Hyper-parameters: PPO} & \textbf{Case study 1} & \textbf{Case study 2}\\
    \hline
    gamma                                                                  &        0.95                    & 0.9                 \\
    learning\_rate                                                         &        0.0007410               & 0.0003843           \\
    nminibatches                                                           &        2                       & 2                   \\
    n\_steps                                                               &        672                     & 256                 \\
    ent\_coef                                                              &        3.141e-03               & 1.226e-06           \\
    cliprange                                                              &        0.3                     & 0.3                 \\
    noptepochs                                                             &        5                       & 10                  \\
    lambda                                                                 &        0.95                    & 0.8                   
    \end{tabular}
    \caption{Best found PPO hyper-parameters}
\end{table}

\begin{table}[!hb]
    \centering
    \begin{tabular}{l|c|c}
    \rowcolor[HTML]{efefef} 
    \textbf{Hyper-parameters: TD3} & \textbf{Case study 1} & \textbf{Case study 2}\\
    \hline
    gamma                                                                  &        0.9               & 0.9                 \\
    learning\_rate                                                         &        7.551e-05         & 0.0003833           \\
    batch\_size                                                            &        24               & 1e2                 \\
    buffer\_size                                                           &        1e5               & 1e5                 \\
    train\_freq                                                            &        96                & 2e3                 \\
    gradient\_steps                                                        &        100               & 2e3                 \\
    noise\_type                                                            &        Ornstein Uhlenbeck & normal               \\
    noise\_std                                                             &        0.337             & 0.329                  
    \end{tabular}
    \caption{Best found TD3 hyper-parameters}
\end{table}

\par The hyper-parameter study of case study 1 was conducted on a local machine with a processor Intel® Core™ i5-9300H CPU @2.4GHz, 8 GB of Ram and an SSD. The total run-time for the study was 17.54 hours (0.73 days) for the PPO and 24 hours (1 day) for the TD3 studies, respectively.

\par The hyper-parameter study of case study 2 was conducted on a virtual cloud instance with 6 vCPU Intel® Xeon® E5-2630 v3 cores, 16 GB of RAM and an SSD. The total run-time for the study was 1175.38 hours (approx. 49 days) for the PPO agent and 1233.89 hours (approx. 51 days) for the TD3 agent.

\par Both case studies have approximately an equal number of trials (50 - 60 trials).

\begin{figure}[H]
\begin{subfigure}{0.45\textwidth}
\includegraphics[width=1\linewidth, height=5cm, trim={0cm 1cm 5cm 3cm}, clip]{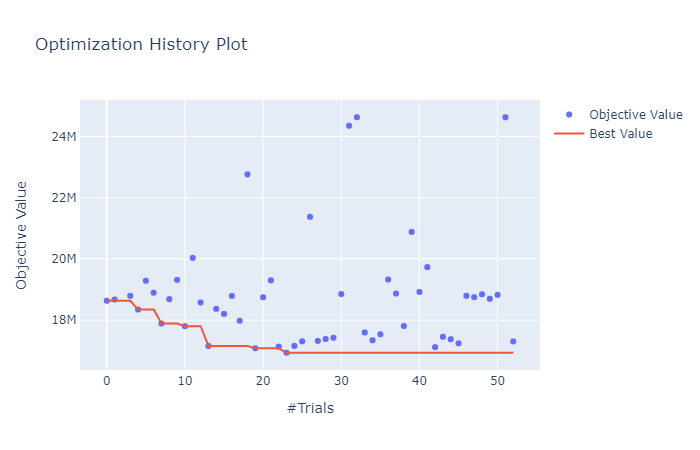} 
\caption{PPO}
\end{subfigure}
\begin{subfigure}{0.45\textwidth}
\includegraphics[width=1\linewidth, height=5cm, trim={0cm 1cm 5cm 3cm}, clip]{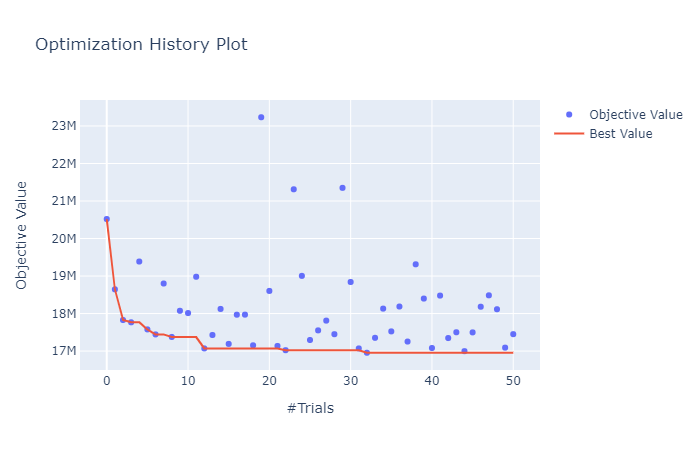}
\caption{TD3}
\end{subfigure}
\caption{Case study I: Hyper-parameter optimization history plots}
\label{fig: optuna_history II}
\end{figure}

\begin{figure}[H]
\begin{subfigure}{0.45\textwidth}
\includegraphics[width=1\linewidth, height=5cm, trim={0cm 1cm 1cm 3cm}, clip]{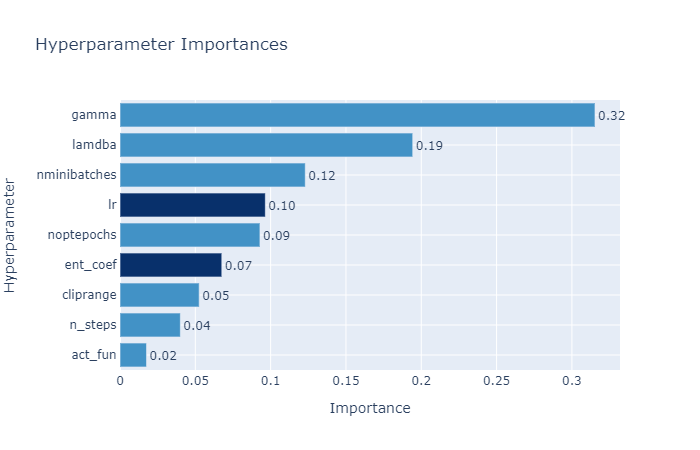}
\caption{PPO}
\end{subfigure}
\begin{subfigure}{0.45\textwidth}
\includegraphics[width=1\linewidth, height=5cm, trim={0cm 1cm 1cm 3cm}, clip]{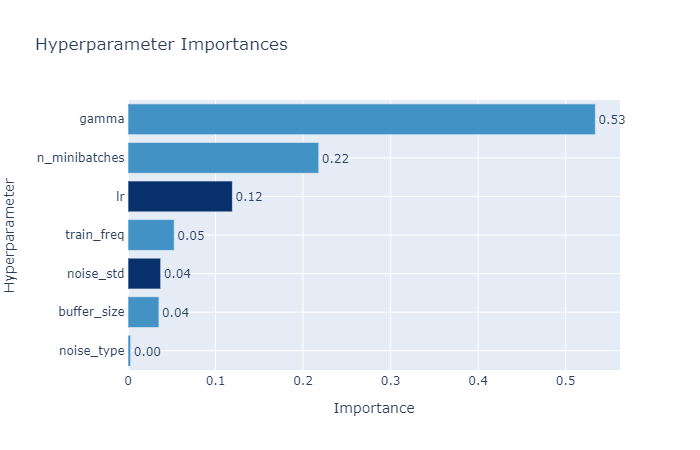}
\caption{TD3}
\end{subfigure}
\caption{Case study I: Hyper-parameter importance plots}
\label{fig: optuna_importance II}
\end{figure}

\begin{figure}[H]
\begin{subfigure}{0.45\textwidth}
\includegraphics[width=1\linewidth, height=5cm]{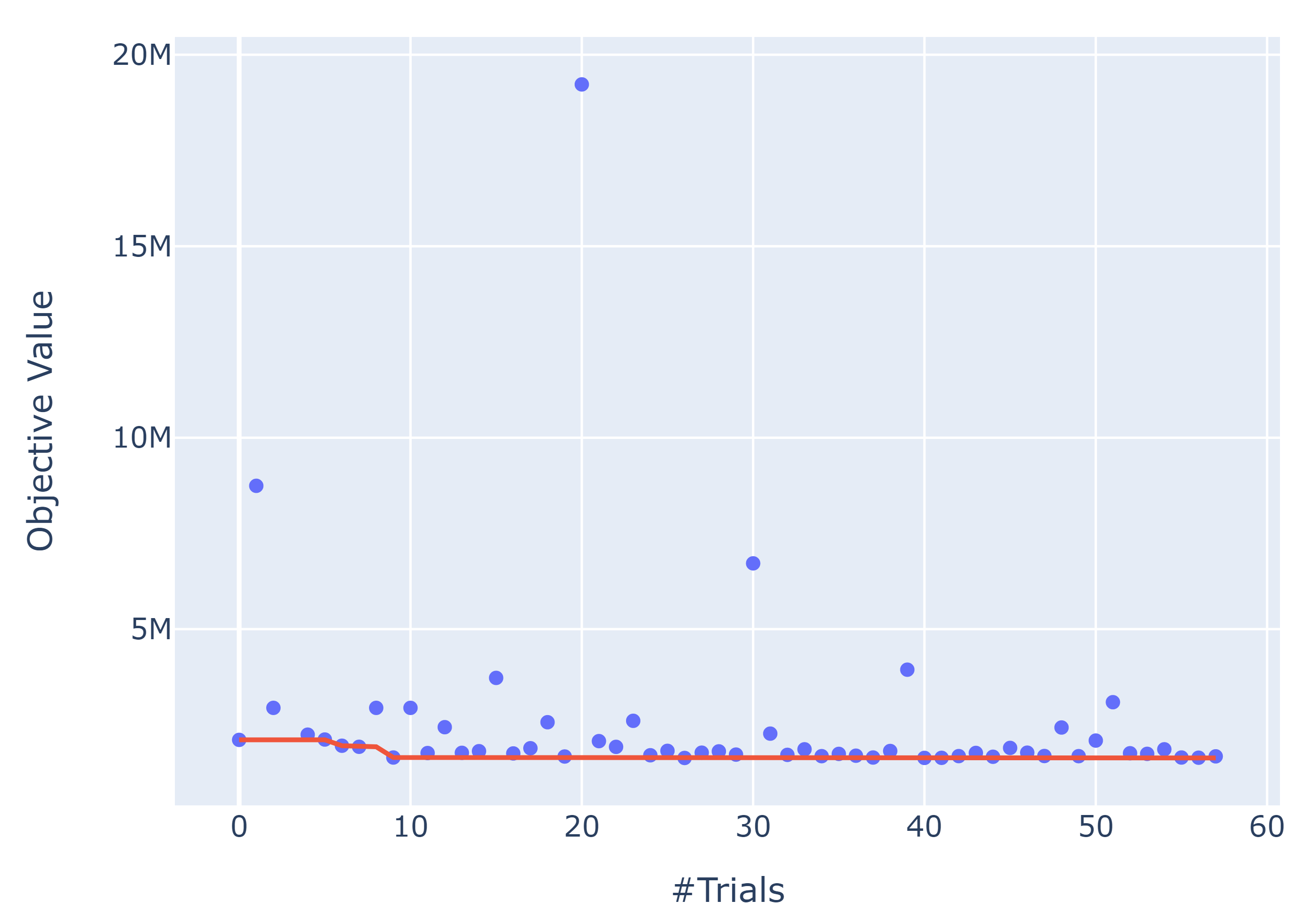} 
\caption{PPO}
\end{subfigure}
\begin{subfigure}{0.45\textwidth}
\includegraphics[width=1\linewidth, height=5cm]{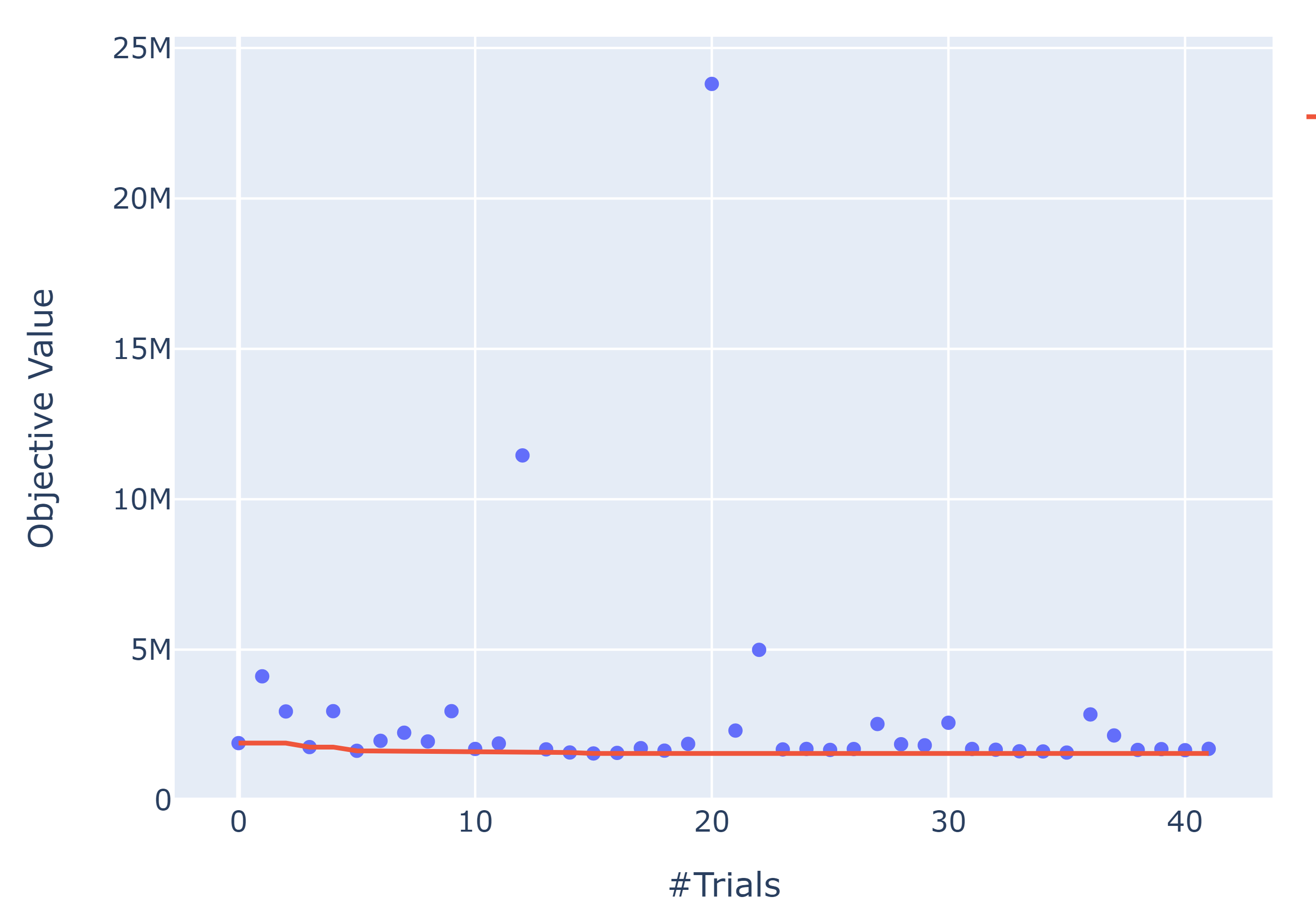}
\caption{TD3}
\end{subfigure}
\caption{Case study II: Hyper-parameter optimisation history plots}
\label{fig: optuna_history}
\end{figure}

\begin{figure}[H]
\begin{subfigure}{0.45\textwidth}
\includegraphics[width=1\linewidth, height=5cm]{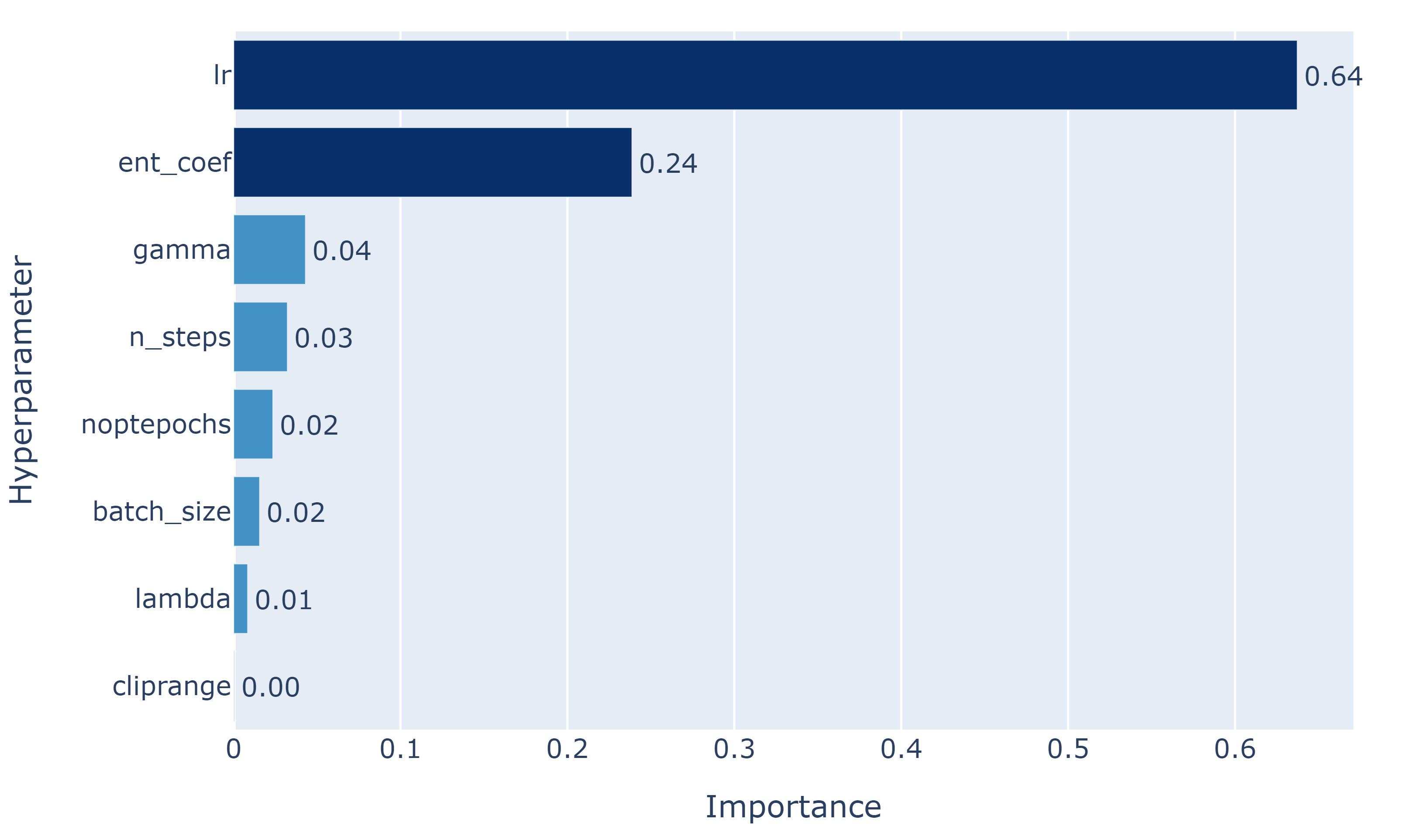} 
\caption{PPO}
\end{subfigure}
\begin{subfigure}{0.45\textwidth}
\includegraphics[width=1\linewidth, height=5cm]{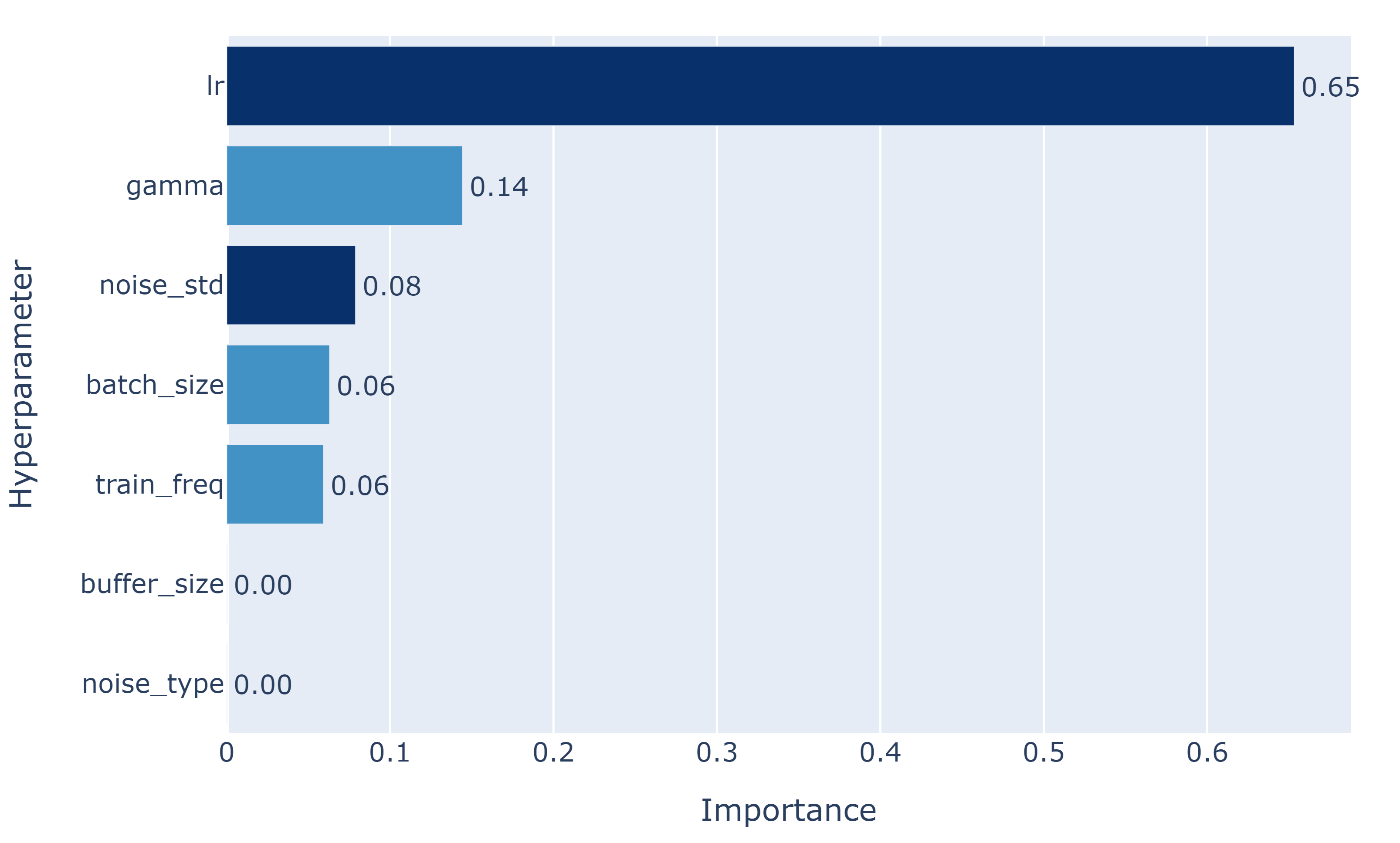}
\caption{TD3}
\end{subfigure}
\caption{Case study II: Hyper-parameter importance plots}
\label{fig: optuna_importance}
\end{figure}

\par The previous figures show relevant information about the hyper-parameter studies performed in each of the case studies. There are two types of figures for each case, one of them depicts the objective value achieved after each trial and the other one the importance of each hyper-parameter involved in the study. The former type of figure (i.e., \autoref{fig: optuna_history II} and \autoref{fig: optuna_history}) shows the best objective value achieved during the study in a red curve. Furthermore, it shows the specific objective value of each trial in the study as blue points. The set of hyper-parameters selected was the first one that achieved the lowest value in the red curve. The latter type of figure (i.e., \autoref{fig: optuna_importance II} and \autoref{fig: optuna_importance}) give an indicator of which hyper-parameter had a greater impact during the study.

\section{Simulations visualization}
\label{Appendix C}
\par This section shows the control policy of TD3 and realistic LMPC for different times during the year (i.e., the whimsical noise of the LMPC is due to its prediction errors). The dates selected correspond to seasons of the year that, due to the weather and market conditions, show a sample of the expected behaviour throughout the year. 
\par The first observation that can be made is the difference between the battery use in the LMPC and the TD3 techniques. While the LMPC takes advantage of the availability of a battery storage system in the three periods, the TD3 agent decides that it is not beneficial to use it. A possible explanation is the selection of hyper-parameters, especially the number of time-steps and discount factors used during the policy update process. These hyper-parameters possibly led the policy towards a local optimum that does not use electrical energy storage. 
\par Additionally, it is possible to see that both techniques resulted in a slight overproduction of thermal energy. Due to the boiler's controller (which aims to produce any missing thermal energy), the thermal discomfort occurred in only one direction: overproduction. This overproduction was caused by the inaccuracy of the model and the forecast predictions in the case of the LMPC and by the lack of an operational constraint in the case of the TD3 agent. However, both techniques have a mechanism to prevent this overproduction from being greater. While the LMPC makes use of the thermal energy balance in its internal model, the TD3 agent minimizes the thermal discomfort penalty that is included in its reward function.
\par 
\begin{figure}[H]
    \centering
    \includegraphics[width=\textwidth,height=\textheight,keepaspectratio]{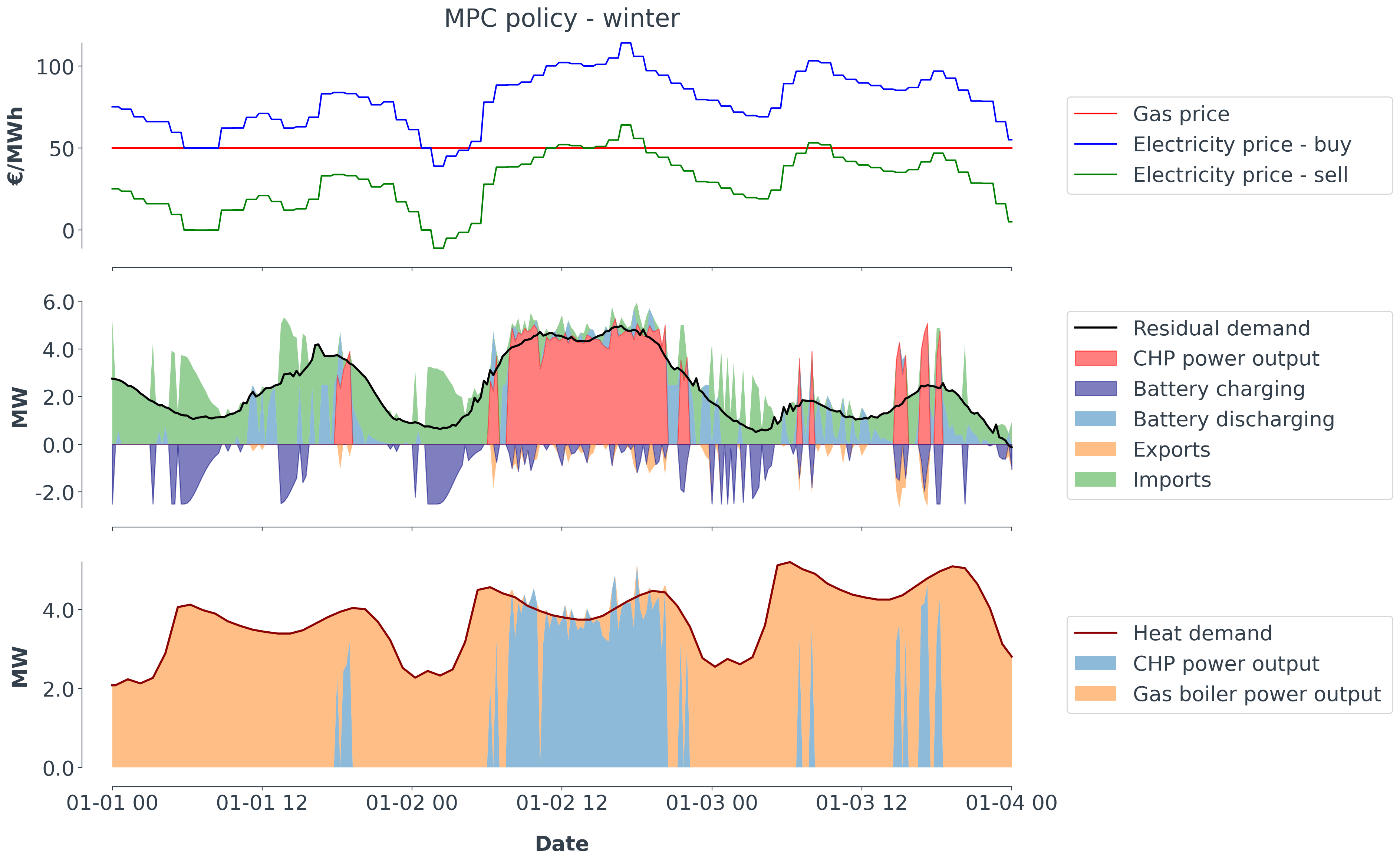}
    \caption{Case study I: LMPC policy - winter}
\end{figure}
\begin{figure}[H]
    \centering
    \includegraphics[width=\textwidth,height=\textheight,keepaspectratio]{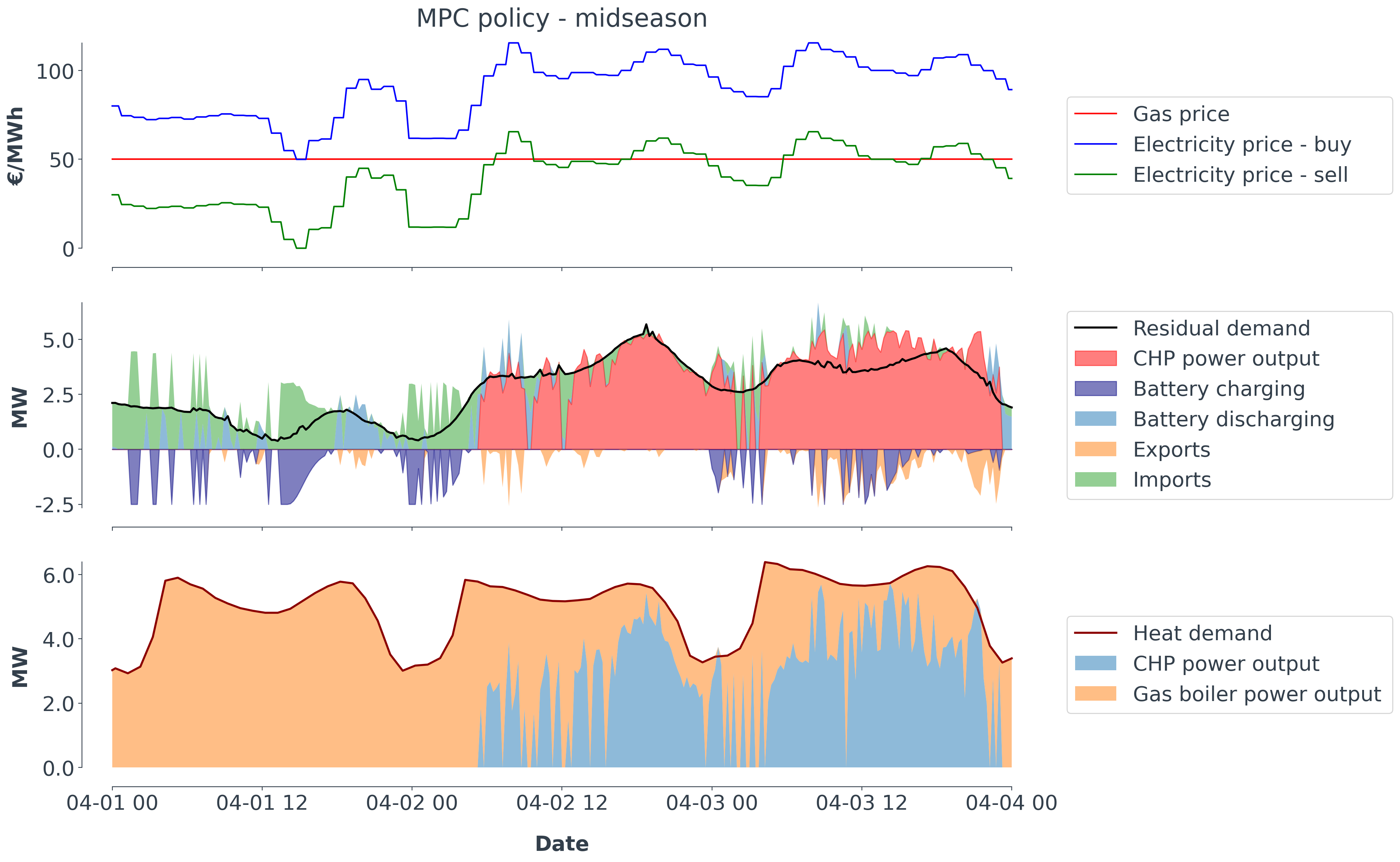}
    \caption{Case study I: LMPC policy - midseason}
\end{figure}
\begin{figure}[H]
    \centering
    \includegraphics[width=\textwidth,height=\textheight,keepaspectratio]{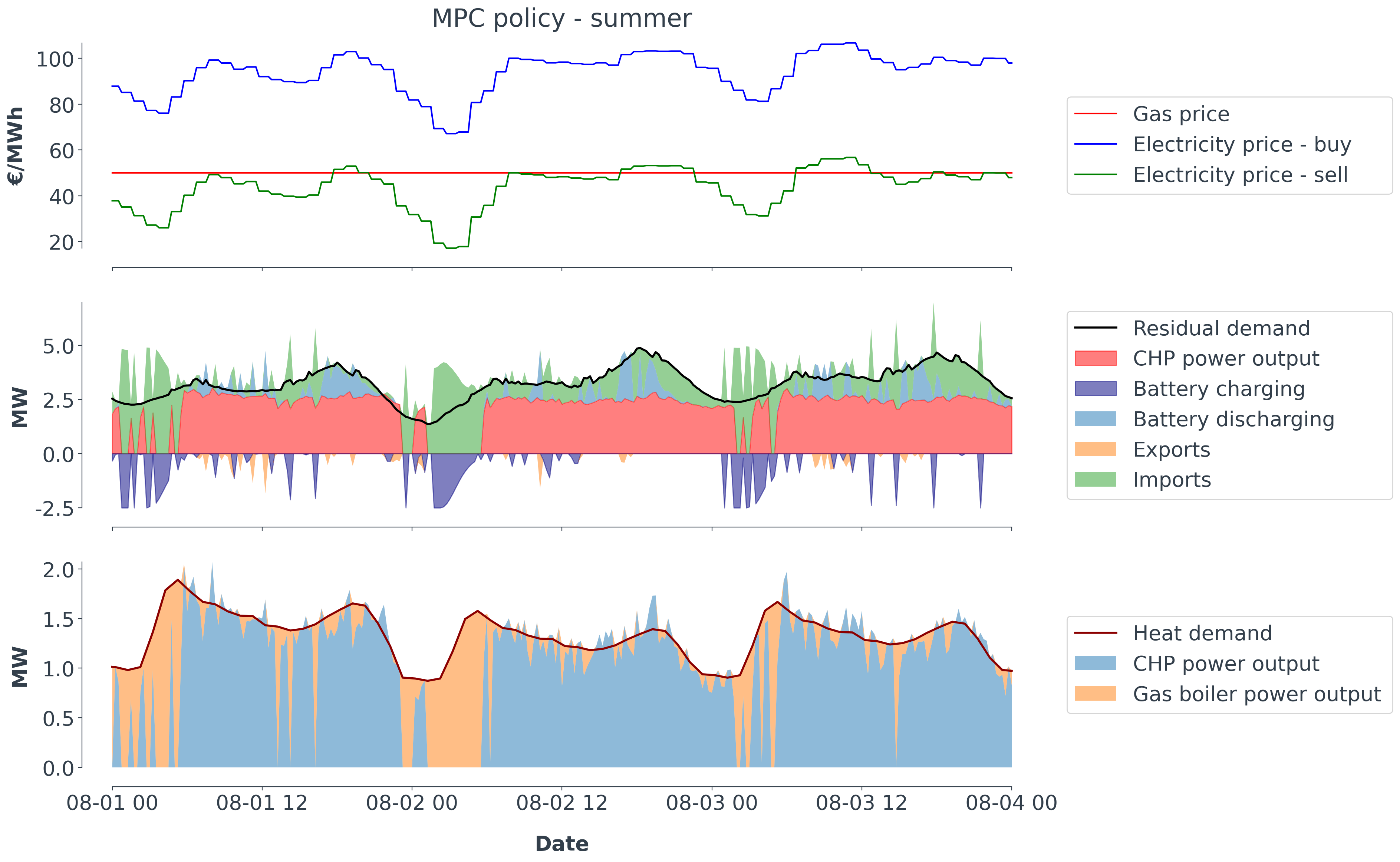}
    \caption{Case study I: LMPC policy - summer}
\end{figure}
\begin{figure}[H]
    \centering
    \includegraphics[width=\textwidth,height=\textheight,keepaspectratio]{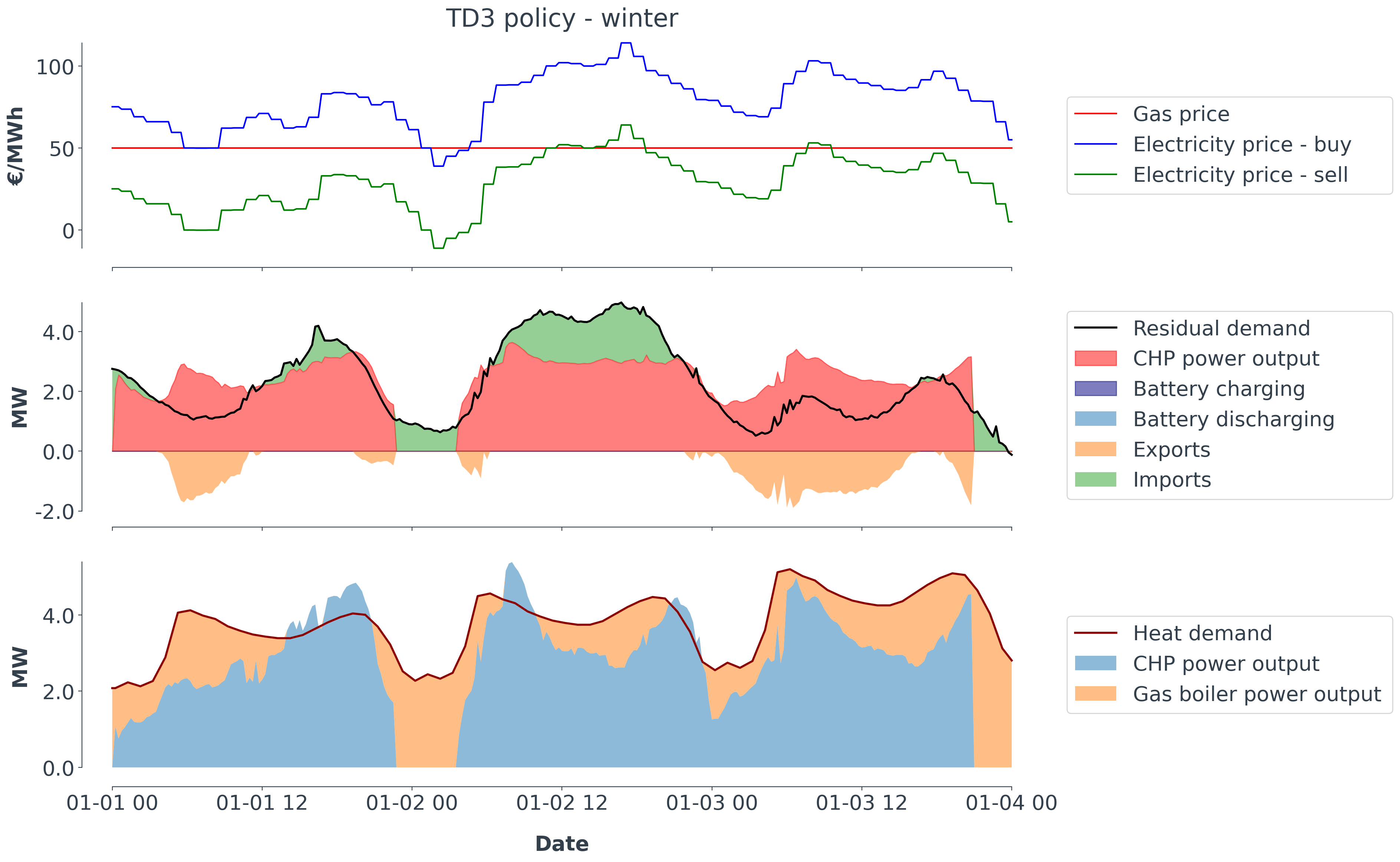}
    \caption{Case study I: TD3 policy - winter}
\end{figure}
\begin{figure}[H]
    \centering
    \includegraphics[width=\textwidth,height=\textheight,keepaspectratio]{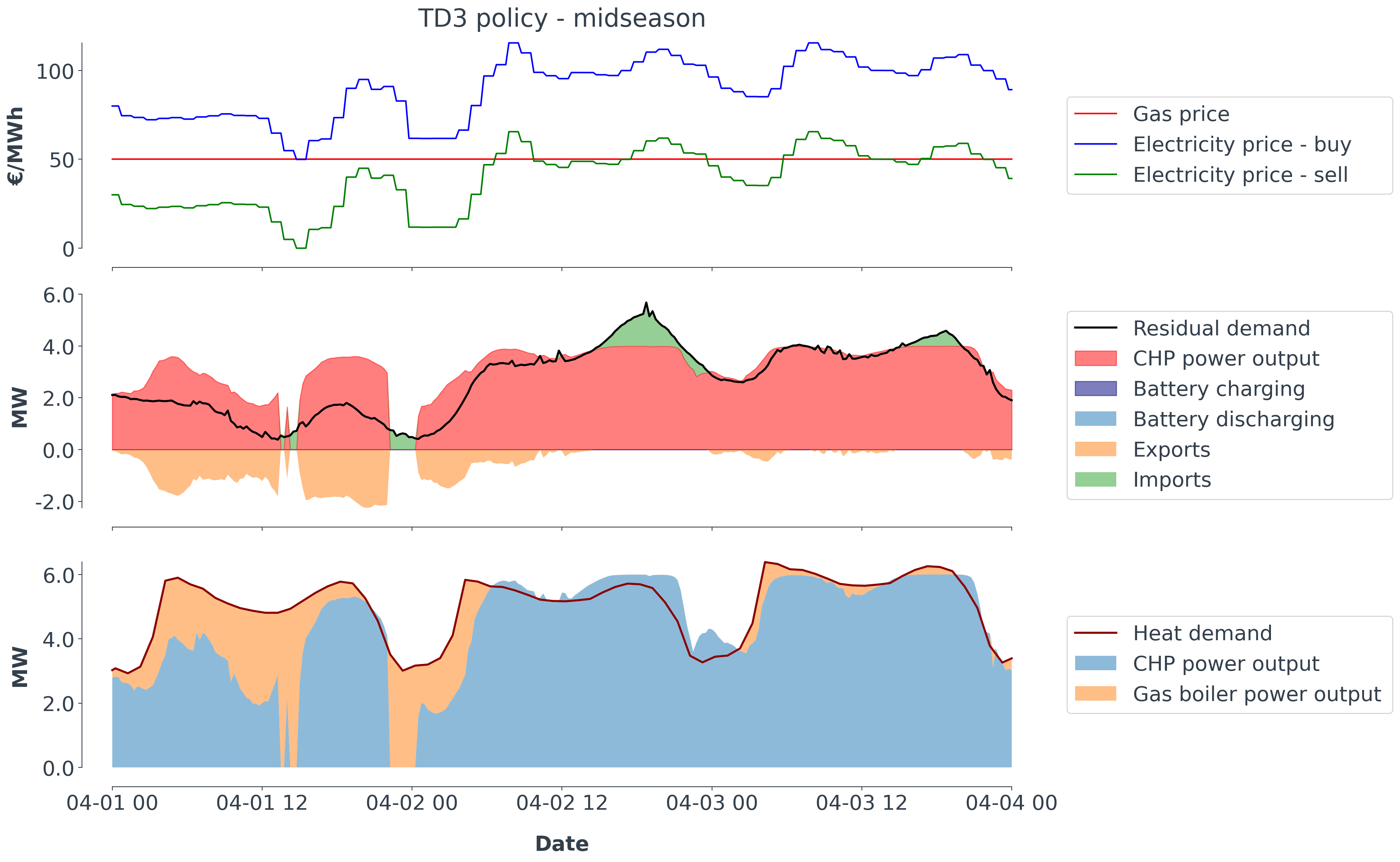}
    \caption{Case study I: TD3 policy - midseason}
\end{figure}
\begin{figure}[H]
    \centering
    \includegraphics[width=\textwidth,height=\textheight,keepaspectratio]{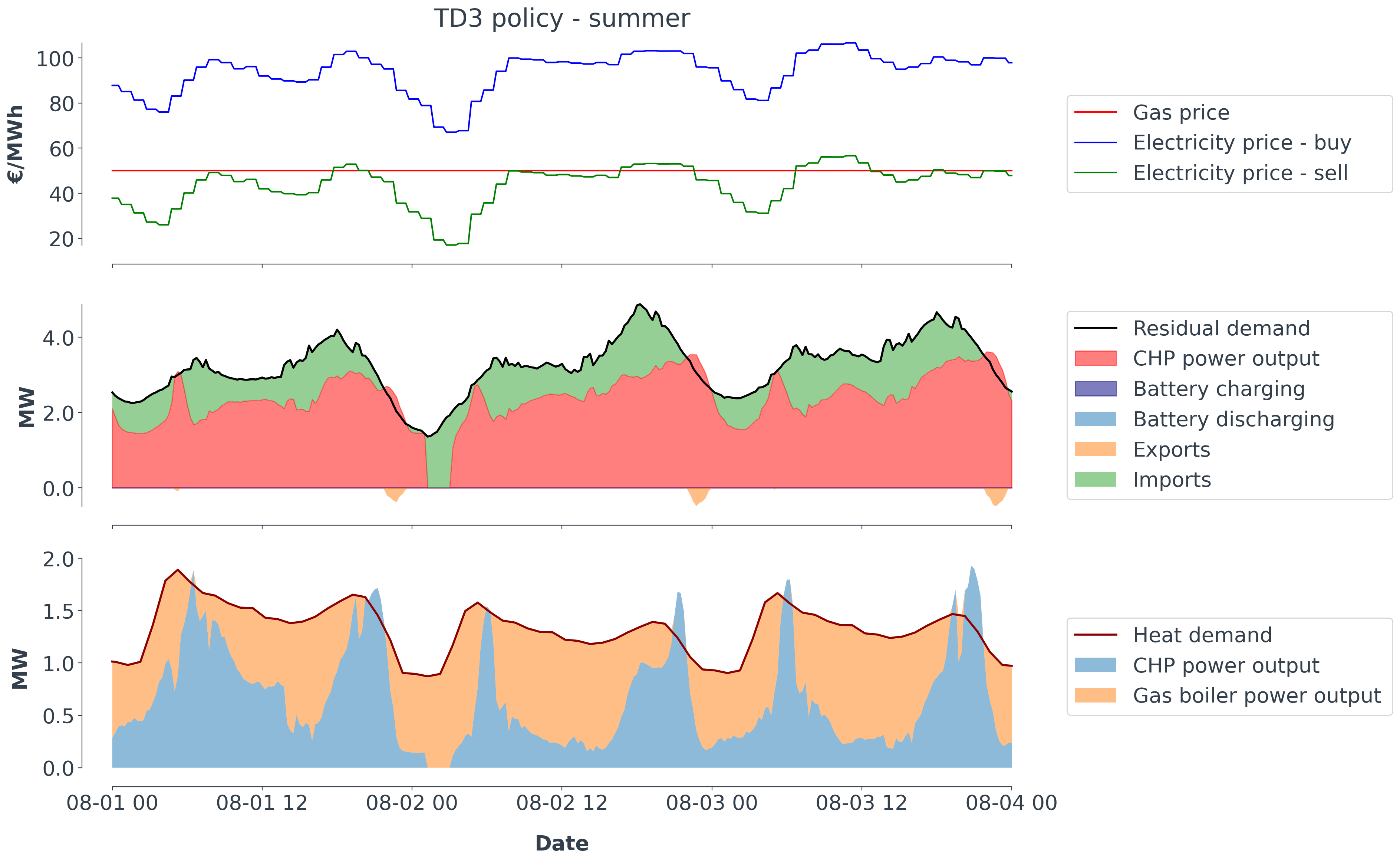}
    \caption{Case study I: TD3 policy - summer}
\end{figure}

\par It is observable in the next three figures that the thermal discomfort was close to zero during winter and slightly higher during the summer. This slight mismatch can be explained by the inaccuracy of the linear model used in the LMPC, since keeping the thermal discomfort equal to zero is a constraint in this control technique, which prevents it to deliberately deviate from zero. The midseason and winter periods show a more diverse use of the multiple assets and an overall reduced thermal discomfort.
\begin{figure}[H]
    \centering
    \includegraphics[width=\textwidth,height=\textheight,keepaspectratio]{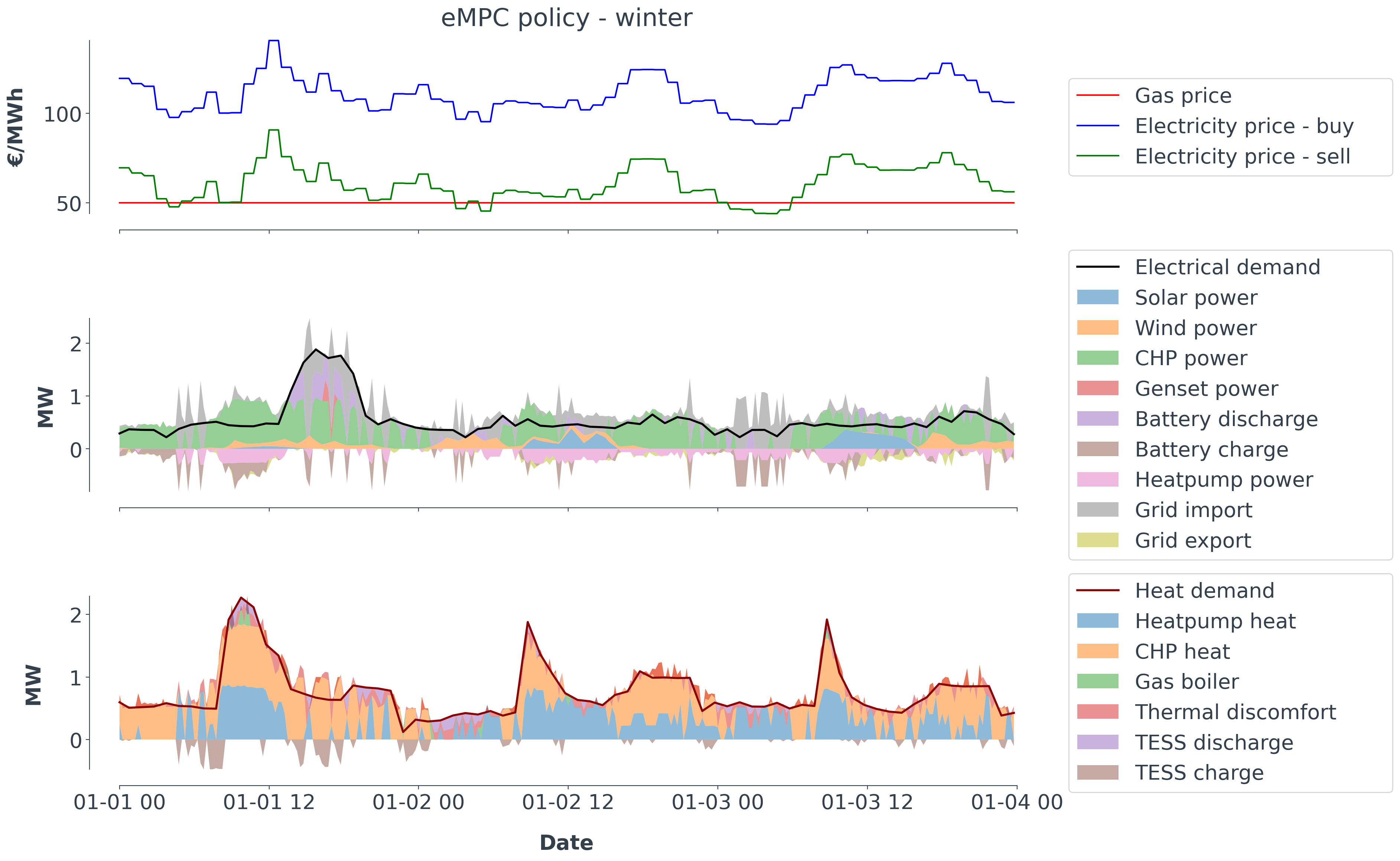}
    \caption{Case study II: LMPC policy - winter}
\end{figure}
\begin{figure}[H]
    \centering
    \includegraphics[width=\textwidth,height=\textheight,keepaspectratio]{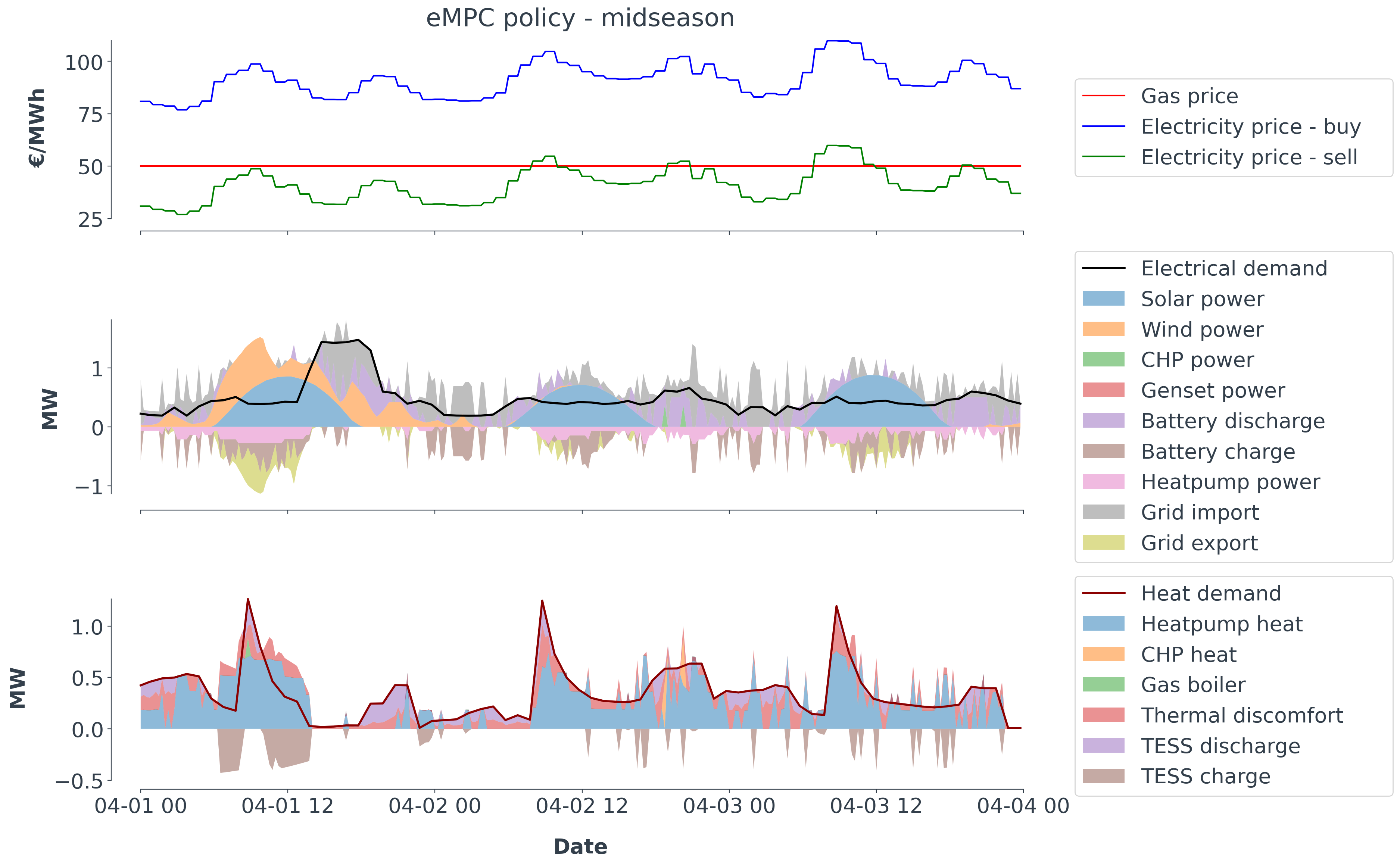}
    \caption{Case study II: LMPC policy - midseason}
\end{figure}
\begin{figure}[H]
    \centering
    \includegraphics[width=\textwidth,height=\textheight,keepaspectratio]{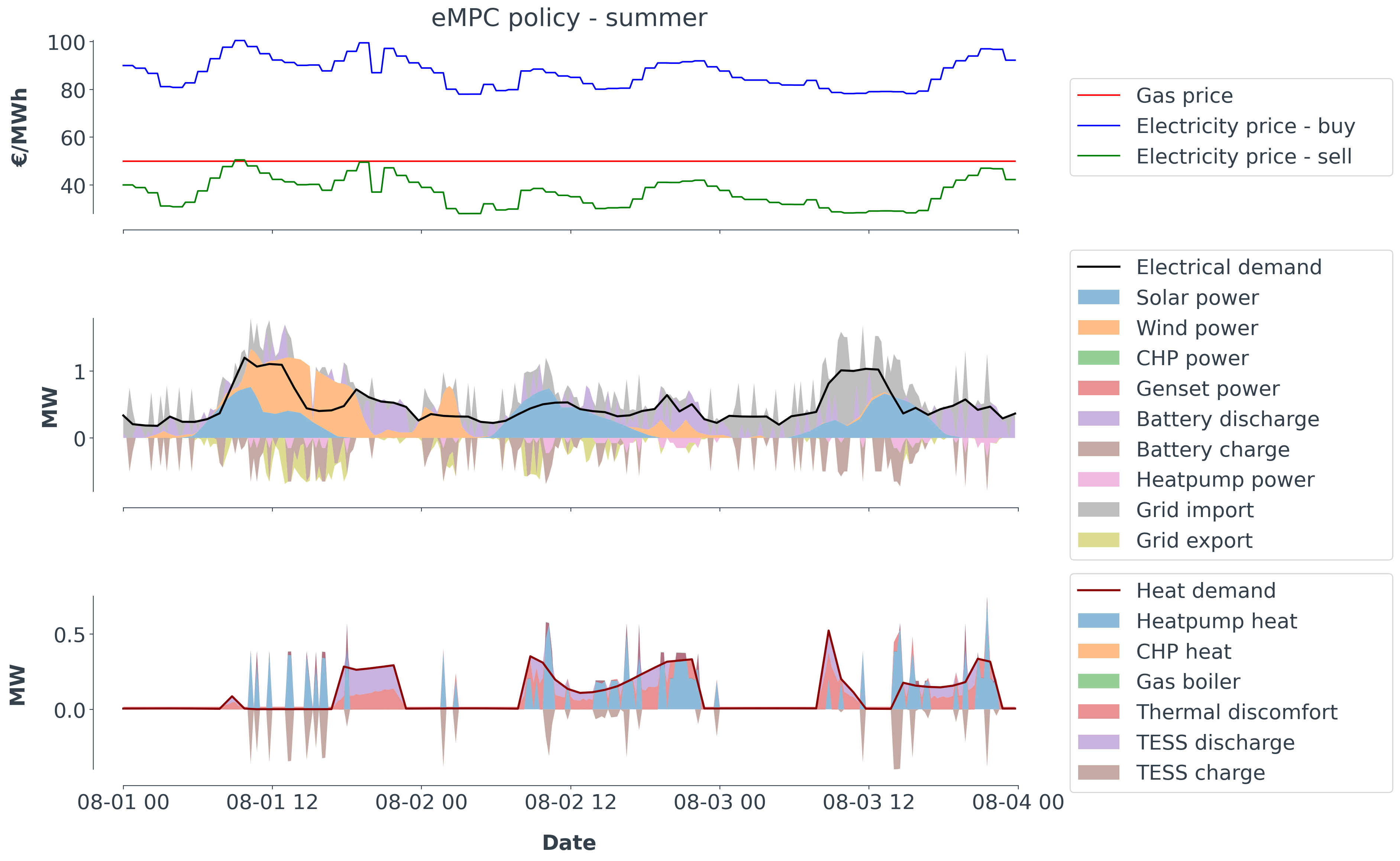}
    \caption{Case study II: LMPC policy - summer}
\end{figure}
\par Similarly to the LMPC, the TD3 agent shows difficulties to avoid thermal discomfort during the summer period. However, the midseason period depicts a lower thermal discomfort than that achieved by the LMPC. In contrast to the previous control technique, the TD3 agent does not have explicit constraints in the thermal comfort, but it uses a term that considers this comfort in the reward function. Therefore, the agent tries to reduce as much as it is convenient according to the reward function and the operational capabilities of the assets.
\begin{figure}[H]
    \centering
    \includegraphics[width=\textwidth,height=\textheight,keepaspectratio]{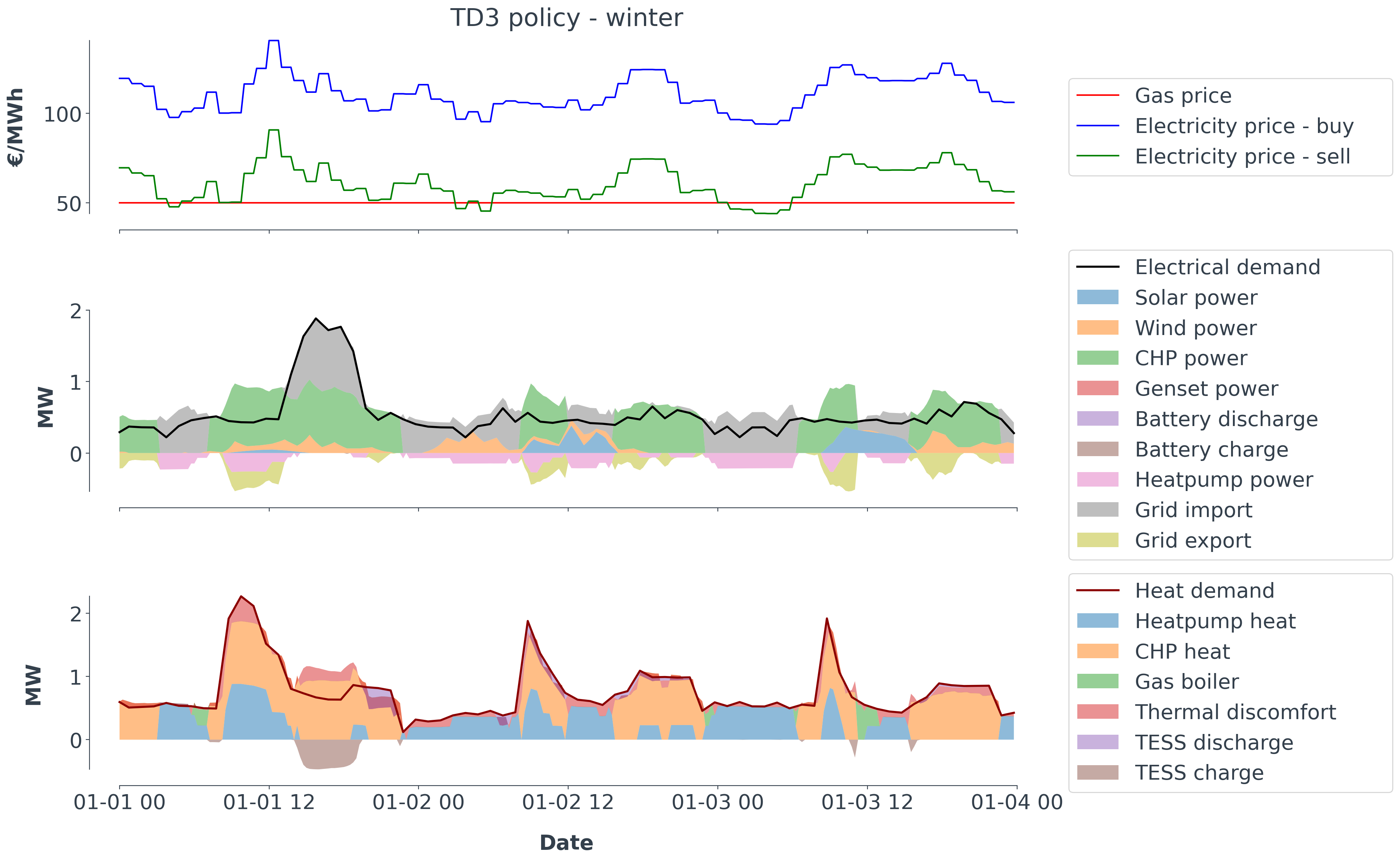}
    \caption{Case study II: TD3 policy - winter}
\end{figure}
\begin{figure}[H]
    \centering
    \includegraphics[width=\textwidth,height=\textheight,keepaspectratio]{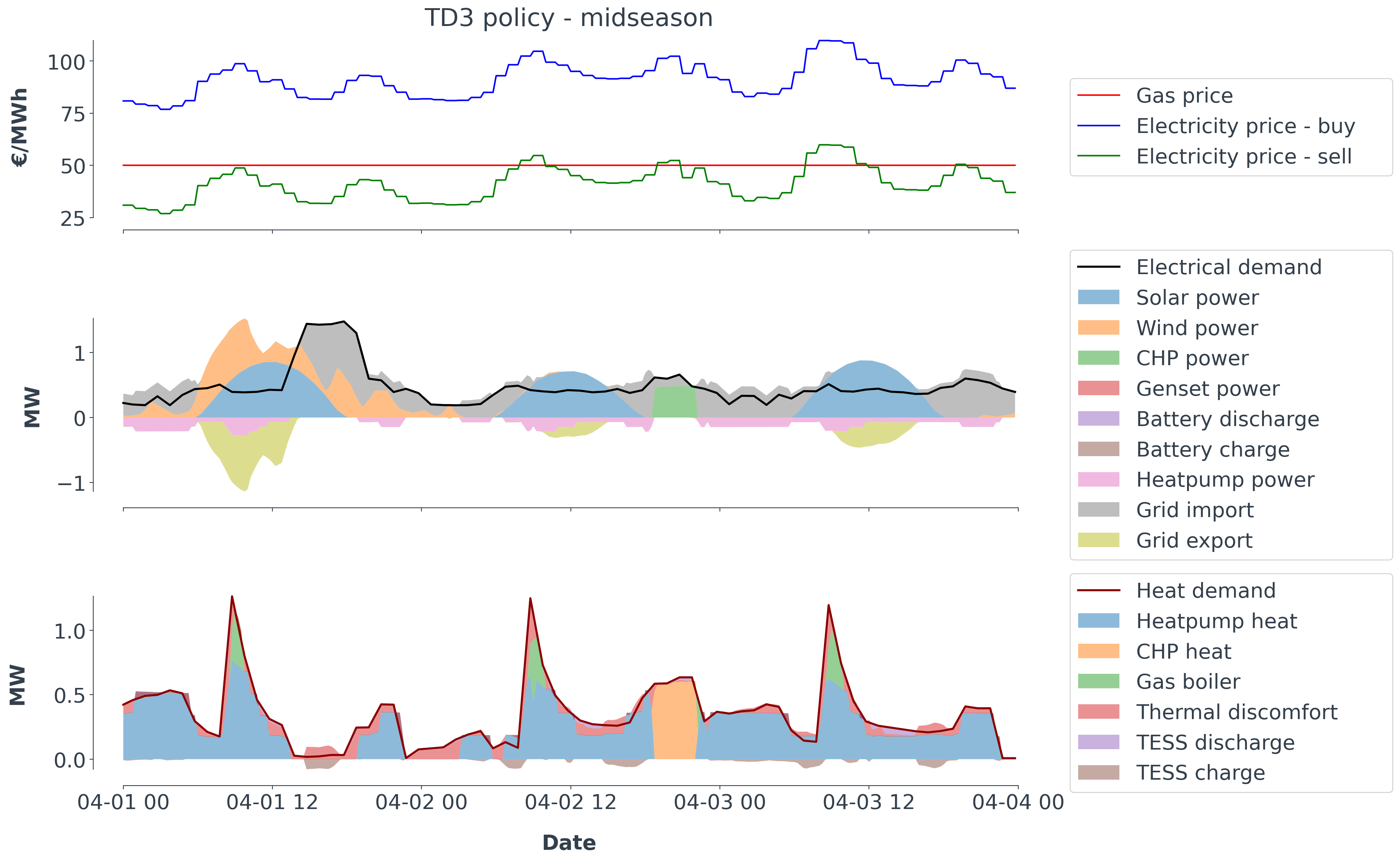}
    \caption{Case study II: TD3 policy - midseason}
\end{figure}
\begin{figure}[H]
    \centering
    \includegraphics[width=\textwidth,height=\textheight,keepaspectratio]{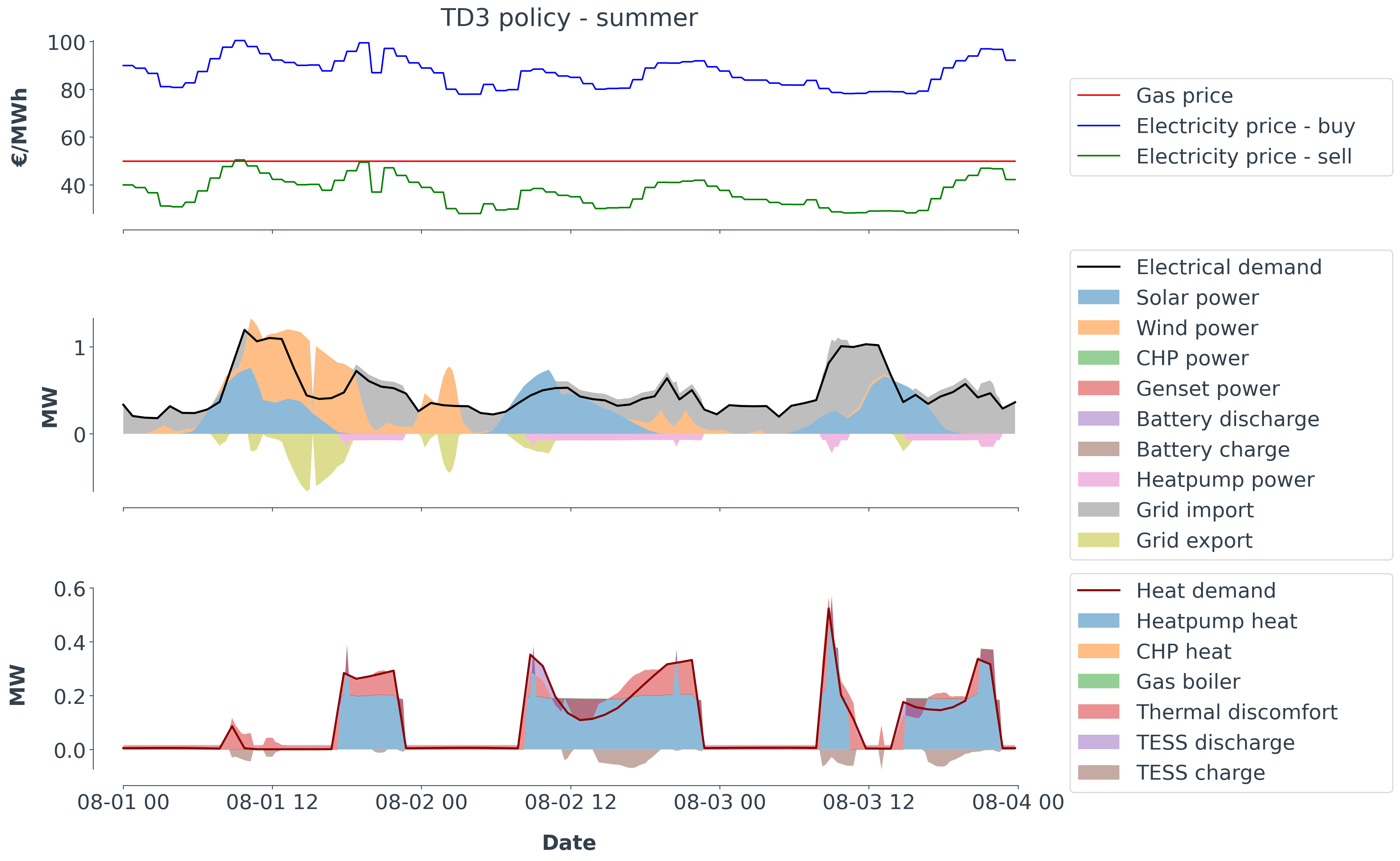}
    \caption{Case study II: TD3 policy - summer}
\end{figure}

\section{Pseudocode of PPO and TD3}
\label{Appendix D}
\begin{algorithm}[H]
\DontPrintSemicolon
\SetAlgoLined
 \nl Input: initial policy parameters \( \theta_0 \), initial value function parameters \( \phi_0 \) \;
 \nl \For{\( k=0,1,2,\dots \)}{
 \nl Collect set of trajectories \( \mathcal{D}_k=\{ \tau_i \} \) by running policy \( \pi_k = \pi(\theta_k) \) in the environment\;
 \nl Compute rewards-to-go \( \hat R_t \)\;
 \nl Compute advantage estimates, \( \hat A_t \) (using any method of advantage estimation) based on the current value function \( V_{\phi_k} \)\;
 \nl Update the policy by maximizing the objective:
 \begin{equation*}
     \theta_{k+1}=\arg \max_\theta \frac{1}{|\mathcal{D}_k|T} \sum_{\tau \in \mathcal{D}_k} \sum_{t=0}^T \min \bigg( \frac{\pi_\theta(a_t|s_t)}{\pi_{\theta_k}(a_t|s_t)} A^{\pi_{\theta_k}}(s_t,a_t), \ g(\epsilon, A^{\pi_{\theta_k}}(s_t,a_t)) \bigg)
 \end{equation*}
 typically via stochastic gradient ascent with Adam.\;
 \nl Fit value function by regression with mean squared error:
 \begin{equation*}
    \phi_{k+1} = \arg \min_\phi \frac{1}{|\mathcal{D}_k|T} \sum_{\tau \in \mathcal{D}_k} \sum_{t=0}^T \big( V_\phi(s_t)-\hat R_t \big)^2
 \end{equation*}
 typically via some gradient descent algorithm.\;
 }
 \caption{PPO-clip \cite{OpenAI2020ProximalDocumentation}}\label{algo: ppo}
\end{algorithm}

\begin{figure}[H]
    \centering
    \includegraphics[width=\textwidth,height=\textheight,keepaspectratio]{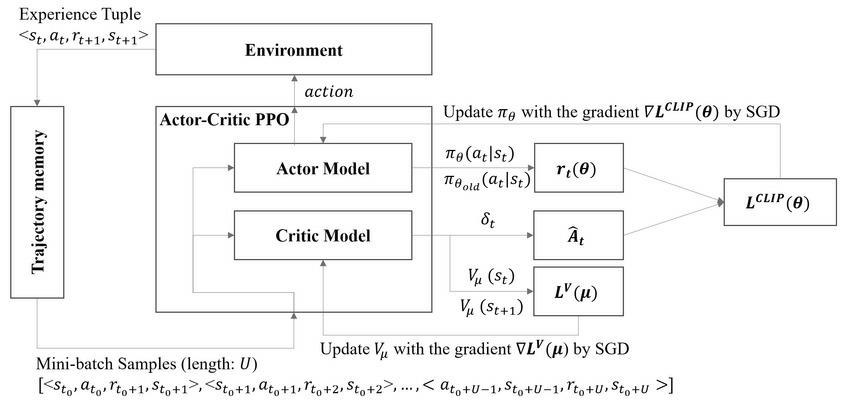}
    \caption{Example flow-chart of the interactions agent-environment in PPO, extracted from \cite{s20051359}. Note that this figure only attempts a generic representation of the learning process described in Algorithm \ref{algo: ppo}. }
\end{figure}

\pagebreak
\begin{algorithm}[H]
\DontPrintSemicolon
\SetAlgoLined
 \nl Input: initial policy parameters \( \theta \), Q-function parameters \( \phi_1 \), \( \phi_2 \), \\ 
 empty replay buffer \(\mathcal{D}\) \;
 \nl Set target parameters equal to main parameters \( \theta_{targ} \leftarrow \theta\), \( \theta_{targ,1} \leftarrow \theta_1\), \( \theta_{targ,2} \leftarrow \theta_2\) \;
 \nl \textbf{repeat} \;
 \nl \hspace{0.2cm} Observe state \(s\) and select action \(a = \text{clip}(\mu_{\theta}(s) + \epsilon, a_{Low}, a_{High}) \), where \( \epsilon \sim \mathcal{N} \) \;
 \nl \hspace{0.2cm} Execute \( a \) in the environment \;
 \nl \hspace{0.2cm} Observe next state \( s' \), reward \( r \) and done signal \( d \) to indicate whether \( s' \) is terminal \;
 \nl \hspace{0.2cm} Store \( (s,a,r,s',d) \) in replay buffer \( \mathcal{D} \) \;
 \nl \hspace{0.2cm} If \( s' \) is terminal, reset environment state \;
 \nl \hspace{0.2cm} \textbf{if} it's time to update \textbf{then} \;
 \nl \hspace{0.4cm} \textbf{for} \( j \) in range(however many updates) \textbf{do} \;
 \nl \hspace{0.6cm} Randomly sample a batch of transitions, \( B = \{ (s,a,r,s',d) \} \) from \( \mathcal{D} \) \;
 \nl \hspace{0.6cm} Compute target actions
 \begin{align*}
     a'(s') = \text{clip}\big( \mu_{\theta_{targ}}(s') + \text{clip}(\epsilon, -c, c), a_{Low}, a_{High} \big),& & \epsilon \sim \mathcal{N}(0,\sigma)
 \end{align*} \vspace{-0.5cm}\;
 \nl \hspace{0.6cm} Compute targets
 \begin{equation*}
     y(r,s',d) = r + \gamma(1 - d) \min_{i=1,2} Q_{\phi_{targ,i}}(s',a'(s'))
 \end{equation*} \vspace{-0.5cm}\;
 \nl \hspace{0.6cm} Update Q-function by one step of gradient descent using
 \begin{align*}
     \nabla_{\phi_{i}} \frac{1}{|B|} \sum_{(s,a,r,s',d)\in B}(Q_{\phi_{i}}(s,a) - y(r,s',d))^2& & \text{for} \ \ i=1,2
 \end{align*} \vspace{-0.5cm}\;
 \nl \hspace{0.6cm} \textbf{if} \( \ j \ \) mod \(\ \texttt{policy\_delay} = 0\) \textbf{then} \;
 \nl \hspace{0.8cm} Update policy by one step of gradient ascent using
 \begin{equation*}
     \nabla_\theta \frac{1}{|B|} \sum_{s\in B} Q_{\phi_{1}}(s,\mu_\theta(s))
 \end{equation*}  \vspace{-0.5cm}\;
 \nl \hspace{0.8cm} Update target networks with
 \begin{align*}
     \phi_{targ,i} \leftarrow \rho \phi_{targ,i} + (1 - \rho) \phi_i& & \text{for} \ \ i=1,2 \\
     \theta_{targ} \leftarrow \rho \theta_{targ} + (1 - \rho) \theta
 \end{align*} \vspace{-0.5cm} \;
 \nl \hspace{0.6cm} \textbf{end if} \;
 \nl \hspace{0.4cm} \textbf{end for} \;
 \nl \hspace{0.2cm} \textbf{end if} \;
 \nl \textbf{until} convergence
 \caption{Twin Delayed DDPG (TD3) \cite{OpenAI2020TwinDocumentation}}\label{algo: td3}
\end{algorithm}

\begin{figure}[H]
    \centering
    \includegraphics[width=\textwidth,height=\textheight,keepaspectratio]{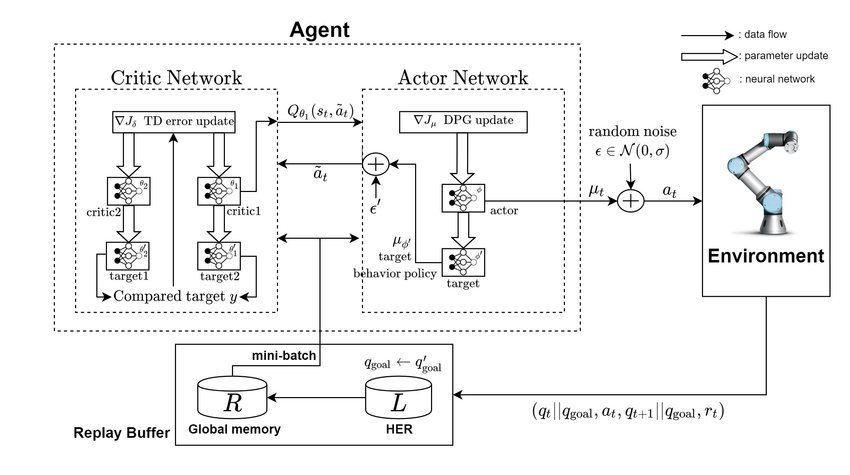}
    \caption{Example flow-chart of the interactions agent-environment in TD3, extracted from \cite{app10020575}. Note that this figure only attempts a generic representation of the learning process described in Algorithm \ref{algo: td3}. }

\end{figure}
\pagebreak

\section{Run-time statistics}
\label{Appendix E}
\par The simulations of case study 1 are conducted on a local machine with a processor Intel® Core™ i5-9300H CPU @2.4GHz, 8 GB of Ram and an SSD. The simulations of case study 2 are conducted on a local machine with a processor Intel® Core™ i5-8365U CPU @1.6GHz, 16 GB of Ram and an SSD. The following run-time statistics, per simulated time-step over a yearly simulation, are observed.

\begin{table}[!htbp]
    \centering
    \begin{tabular}{l|c|c|c|c|c}
    \rowcolor[HTML]{efefef} 
    \textbf{Optimal controller} & \textbf{min} & \textbf{mean} & \textbf{std} & \textbf{max} & \textbf{total}\\
    \hline
    LMPC - perfect predictions & 3.51 s & 10.92 s & 1.20 s & 29.59 s & 382700.55 s \\
    LMPC - realistic predictions & 3.75 s & 10.80 s & 0.5142 s & 12.45 s & 378434.74 s \\
    PPO agent - policy execution & 0.0010 s & 0.0031 s & 0.0017 s & 0.2139 s & 107.93 (+3364) s \\ 
    TD3 agent - policy execution & 0.0009 s & 0.0031 s & 0.0014 s & 0.1591 s & 107.62 (+3685) s
    \end{tabular}
    \caption{Run-time statistics of case study 1}
    \label{tab: run-time 1}
\end{table}

\begin{table}[!htbp]
    \centering
    \begin{tabular}{l|c|c|c|c|c}
    \rowcolor[HTML]{efefef} 
    \textbf{Optimal controller} & \textbf{min} & \textbf{mean} & \textbf{std} & \textbf{max} & \textbf{total}\\
    \hline
    LMPC - perfect predictions & 0.0007 s & 0.1397 s & 3.20 s & 452.85 s & 4897.03 s   \\
    LMPC - realistic predictions & 0.0007 s & 0.1116 s & 1.15 s & 62.13 s & 3910.51 s \\
    PPO agent - policy execution & 0.0013 s  & 0.0032 s & 0.0062 s & 0.9820 s & 112.57 (+1561) s \\
    TD3 agent - policy execution & 0.0010 s  & 0.0042 s & 0.0040 s & 0.0785 s & 147.89 (+2856) s
    \end{tabular}
    \caption{Run-time statistics of case study 2}
    \label{tab: run-time 2}
\end{table}

\par Notice that the mean run-time per time-step is orders of magnitude faster than the control horizon of 15 minutes (i.e. when \(C = T_s\)), and that the maximum run-time per time-step never exceeds this control horizon (as it otherwise would be impractical). However, the LMPC has a significantly longer run-time compared to the RL agents. This longer run-time is due to the LMPC needing to solve a mixed-integer linear program (MILP) with every new receding control horizon, whereas the RL agents simply can execute the policy found up to that point (i.e. "policy execution" statistics) and update its function approximation algorithms (in our case multi-layer perceptrons). 
\par The total training run-time (for 250,000 time-steps) is added between brackets to the total policy execution run-time of the RL agents in \autoref{tab: run-time 1} and \autoref{tab: run-time 2}. Assuming an offline training approach (i.e. pre-training), this total training run-time would be required \textit{a priori}. Assuming an online training approach, the mean run-time for case study 1 would result in 0.0135 s for the PPO agent and 0.0147 s for the TD3 agent and for case study 2 in 0.0062 s for the PPO agent and 0.0114 s for the TD3 agent - which shows the required computational power \textit{including} the function approximation update step (i.e. fitting of multi-layer perceptrons).


\reftitle{References}

\externalbibliography{yes}
\bibliography{references, references_002}



\end{document}